\documentclass[aps,prl,superscriptaddress,twocolumn]{revtex4-2}

\usepackage{graphicx}
\usepackage{amsmath}

\begin{document}

\title{Single-stage few-cycle pulse amplification}

\author{Sagnik Ghosh}
\affiliation{Dept. of Electronics and Communication Engineering, Manipal Institute of Technology, Udupi-Karkala Rd, Eshwar Nagar, Manipal, Karnataka 576104 India}
\author{Nathan G. Drouillard}
\affiliation{Dept. of Physics, University of Windsor, Windsor ON N9B 3P4 Canada}
\author{TJ Hammond}
\email{thammond@uwindsor.ca}
\affiliation{Dept. of Physics, University of Windsor, Windsor ON N9B 3P4 Canada}

\date{\today}

\begin{abstract}
Kerr instability can be exploited to amplify visible, near-, and mid-infrared ultrashort pulses. We use the results of Kerr instability amplification theory to inform our simulations amplifying few-cycle pulses. We show that the amplification angle-dependence is simplified to the phase matching condition of four-wave mixing when the intense pump is considered. Seeding with few-cycle pulses near the pump leads to broadband amplification without spatial chirp, while longer pulses undergo compression through amplification. Pumping in the mid-IR leads to multi-octave spanning amplified pulses with single-cycle duration not previously predicted. We discuss limitations of the amplification process and optimizing pump and seed parameters to maximize amplification.
\end{abstract}


\maketitle

\section{Introduction}

Creating intense few-cycle pulses is a complex process but they are fundamental to attosecond science \cite{CalegariJPB2016}. Laser amplifiers are constrained to a few tens of femtoseconds (1~fs = $10^{-15}$~s) due to gain narrowing \cite{SeresOL2003}, which occurs when the gain medium emission bandwidth narrows the spectrum with each pass, although techniques for spectral control have been developed to improve the pulse bandwidth and phase \cite{WeinerOC2011, LiuOptics2020}. A more versatile and tunable system is the optical parametric amplifier (OPA), which uses the $\chi^{(2)}$ nonlinear susceptibility and has become the source for next-generation attosecond experiments\cite{KobayashiJPB2012, FattahiOptica2014}. These experiments include high harmonic generation cutoff extension \cite{PopmintchevScience2012}, high harmonic generation in condensed matter \cite{GhimireNP2011, VampaNature2015}, and ultrafast pump-probe experiments \cite{GallmannARP2012, NisoliCR2017, LiNC2020}. The nonlinear crystals central to the parametric gain can limit the amplified bandwidth, but non-collinear phase matching can reduce this effect. In either the laser or OPA case, the amplified pulse is focused into a Kerr $\chi^{(3)}$ nonlinear medium to broaden the spectrum followed by pulse compression to create the few-cycle pulse \cite{NisoliOL1997, LuOptica2014}.

Despite the versatility of $\chi^{(2)}$ OPAs, not every material has a second order response, but all materials exhibit Kerr nonlinearity. Recently, exploiting this nonlinearity through Kerr instability amplification (KIA) to directly amplify ultrashort pulses in the visible and near infrared (IR) has been demonstrated \cite{VampaScience2018, JACOL2021}. While four-wave mixing (FWM) also uses the Kerr effect and has been used to amplify a $\sim 75$~nm broad spectrum, this spectral region was far from \--- with no substantive gain near \--- the pump wavelength \cite{ValtnaOL2008}. Conversely, KIA predicts a continuous spectrum of high gain about the pump wavelength, extending over an octave. The large gain and broad bandwidth predicted by KIA makes it a viable candidate as a simplified alternative for single-stage few-cycle pulse amplification. 

In this work, we explore the possibility of few-cycle pulse amplification by KIA by comparing its theory to simulations, the parameters to optimize the amplification, and the limitations of this process.

\section{Kerr instability amplification and pulse propagation}

The theory of KIA has been derived elsewhere \cite{NesrallahOptica2018, Nesrallahbook}, with the necessary equations to reproduce the KIA results presented in Ghosh \textit{et al} \cite{GhoshXXX}. Here, extending FWM, we derive a simplified non-collinear angle criterion for amplification and give an expression for a frequency-dependent gain to reproduce the KIA results.

The conservation of energy in FWM requires that $\omega_p+\omega_p = \omega_s + \omega_i$ where $\omega_{p,s,i}$ are the pump, signal, and idler frequencies, respectively \cite{BloembergenJOSA1980}. The conservation of momentum imposes the phase matching condition $\mathbf{k}_p+\mathbf{k}_p = \mathbf{k}_s + \mathbf{k}_i$ where $\mathbf{k}_{p,s,i}$ are the pump, signal, and idler wave vectors, respectively. With the relation $k = n(\omega)\omega/c$, the phase matching condition of FWM requires that the signal is noncollinear, which is shown with the magenta dotted curve in Fig 1(a) \cite{DubietisLC2008, KobayashiIEEE2012}. We compare the FWM result with KIA, where the criterion for maximum KIA gain is shown in the solid black curve. Although the two curves follow a similar shape, a significant difference arises near the pump wavelength \cite{RubinoOL2011}. This difference arises from the nonlinear index of refraction of the material.

\begin{figure}[h]
\includegraphics[width=0.95\columnwidth]{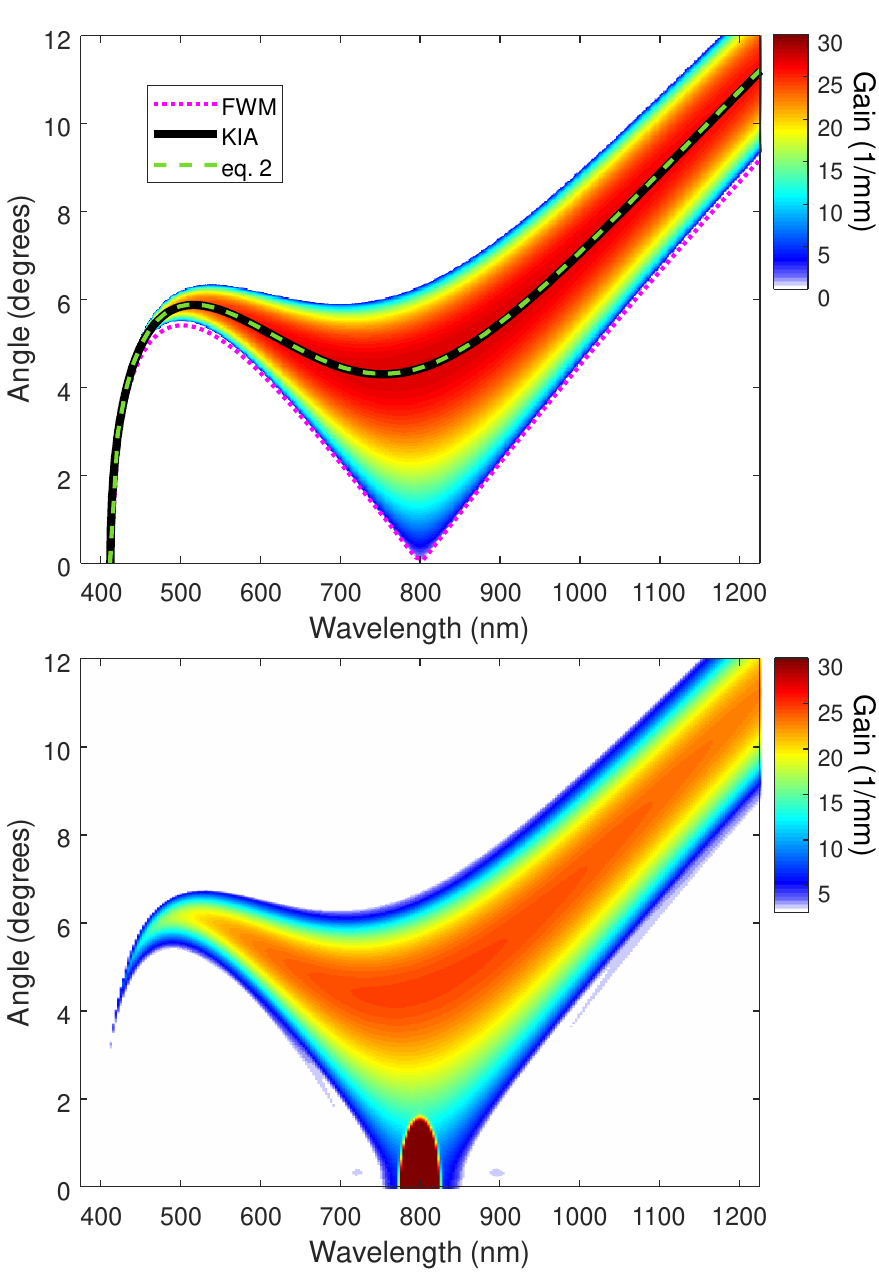}
\caption{Simulation of angle and wavelength amplification dependence of KIA in MgO, with pump wavelength at 800~nm and peak intensity $1.5\times10^{13}$~W/cm$^2$. (a) Gain calculated from KIA theory in the plane-wave single-frequency regime. The phase matching condition of FWM (dotted magenta) disagrees with the optimum external angle for maximum gain predicted by KIA (solid black curve); modification to the phase matching accounting for the nonlinear index leads to good agreement (dashed green). (b) Simulated gain (see text for details) showing angle and spectral changes caused by pump pulse propagation; pump pulse duration of 800~fs and waist of 400 \textmu m leads to the saturated shape for small angles near 800~nm. Crystal thickness is 0.5~mm.}
\end{figure}

We modify the FWM phase matching condition using the definition of the pump wave vector given in Nesrallah \textit{et al} \cite{NesrallahOptica2018},
\begin{equation}
k_p^2 = \frac{\omega_p^2}{c^2}\Bigl(n_p^2 + n_2 I_p \Bigr), \label{eq:pumpwave}
\end{equation}
where $n_p=n(\omega_p)$ is the index at the pump frequency and $n_2$ is the nonlinear Kerr coefficient of the gain medium, and $I_p$ is the peak pump intensity. With this modification, the phase matching condition leads to an (internal) output angle of the signal,
\begin{equation}
\cos\theta_s = \frac{4 (n_p^2 - n_2 I_p) \omega_p^2 + n_s^2 \omega_s^2 - n_i^2 \omega_i^2}{4 n_s \omega_s \omega_p \sqrt{n_p^2 - n_2 I_p}}, \label{eq:phasecurve}
\end{equation}
where $n_{s,i} = n(\omega_{s,i})$ are the indices of refraction of the signal and idler, respectively. The vacuum output angle is determined by Snell’s law, $\theta_f = \sin^{-1}\Bigl(n_s \sin\theta_s \Bigr)$. 

We compare the angle-dependence of optimum gain from KIA and FWM in Fig. 1(a). The angle is the relative (vacuum) angle between the amplified and pump beams in the plane wave limit. There are three regions that show distinct angle dependencies about the pump wavelength, $\lambda_p = 800$~nm: $\lambda \sim \lambda_p/2$, $\lambda \sim \lambda_p$, and $\lambda > \lambda_p$. In the first region, the angle that maximizes the amplification rapidly changes as a function of wavelength. In the second region, from 500~nm to 1000~nm, we see that there is a broad bandwidth region where the angle dependence is nearly constant. Beyond 1000~nm, the angle of optimum gain is linearly dependent on wavelength, continuing for several microns. 

The relative pump-seed angle for maximum gain in KIA (solid black) and eq. \ref{eq:phasecurve} (dashed green) agree, offering a reinterpretation of the KIA angle dependence. Rather than the phase matching of KIA being automatically satisfied \cite{NesrallahOptica2018}, the phase matching condition becomes intensity dependent. The KIA angle dependence shows that there is a broad region that can amplify efficiently, with a maximum gain, $g \approx$ 27/mm. The effect of the pump intensity becomes apparent near the pump wavelength as the value of $n_2 I_p$ becomes large. As the intensity increases, the intensity-dependent phase matching angle increases near the pump and creates a region where a supercontinuum spectrum can be amplified with the same pump-seed angle. Because this change is pump intensity dependent operating near the damage threshold for MgO, this effect is not commonly observed in materials with high $n_2$. Although other materials have higher $n_2$, they damage at this high pump intensity, and thus would not amplify the supercontinuum. We have found that MgO has one of the highest damage thresholds, making it useful for KIA and high harmonic generation from condensed matter \cite{JACOL2021, YouNatPhys2016, KorobenkoOptExp2019, KoJPB2020}. We note that without an intensity-dependent index, FWM phase matching condition  does not predict any useful amplification near the pump wavelength \cite{ValtnaOL2008}. 

We note that the maximum gain occurs at the pump frequency $\omega_p$, and falls off symmetrically in $\omega$. The gain can be approximated about $\omega_p$ by
\begin{equation}
g \approx \frac{\omega_p}{c} \frac{n_2}{n_p} I_p \bigg(2 - \cosh\Big(\frac{\omega - \omega_p}{\omega_p}\Big) \bigg). \label{eq:gainest}
\end{equation}
This gain function agrees with KIA over the region where phase matching occurs, $0 \lesssim \omega_p \lesssim 2\omega_p$, although it slightly overestimates the gain near these extreme regions. The amplification factor is then $A = e^{g L}$, where $L$ is the nonlinear medium length, assuming that the seed and pump are well overlapped throughout the medium. We note that $n_p$ in the denominator could have important implications when $n_p \rightarrow 0$ such as in epsilon-near-zero materials \cite{AlamScience2016, ReshefNatMat2019}.

In the remaining figures, we simulate the amplification of broadband pulses in MgO in two dimensions. We solve the scalar equation of the forward Maxwell’s equation (FME) \cite{HousakouPRL2001, BergePRA2013}, shown to accurately simulate intense single-cycle pulses with multi-octave spanning spectra \cite{HammondOptica2017},
\begin{align}
\frac{\partial}{\partial z} E(\mathbf{r},\omega) &= i\frac{c}{2 n(\omega) \omega} \nabla_{\perp}^2 E(\mathbf{r},\omega) \nonumber \\
&+ i \frac{\omega}{c} \Bigl( n(\omega) - n_g \Bigr) E(\mathbf{r}, \omega) \nonumber \\
&+  i \frac{\mu_0 \omega c}{2 n(\omega)} P_{NL}(\mathbf{r}, \omega) \label{eq:FME}
\end{align}
where, for 2D propagation, $\nabla_{\perp}^2 = \frac{\partial^2}{\partial x^2}$ is the transverse coordinate. We only account for the instantaneous $\chi^{(3)}$ nonlinear polarizability, $P_{NL}= D \epsilon_0 \chi^{(3)} |E|^2 E(x,t)$ and linear polarization; we use the prefactor $D = 3/8$ \cite{AgrawalNFO}. FME takes into account the frequency dependent index of refraction, $n(\omega)$ \cite{Stephens1952} (where $n_g$ is the group index evaluated at the seed central wavelength), although we assume a wavelength (and time) independent nonlinear Kerr coefficient $n_2 = 3 \chi^{(3)} /[4 \epsilon_0 c n_p^2] = 4 \times 10^{-16}$~cm$^2$/W \cite{AdairPRB1989}. 

We note that to obtain agreement with KIA in our simulations, we scale the nonlinear Kerr coefficient by the linear index. The perturbation wave vector is \cite{NesrallahOptica2018},
\begin{align}
k_{\nu} &= \frac{\omega}{c} n(\omega) \sqrt{1 + 2\frac{n_2 I_p}{n^2(\omega)}}  \\
&\approx \frac{\omega}{c} n(\omega) + \frac{\omega}{c} \frac{n_2 I_p}{n(\omega)}.
\end{align} 
Due to this definition, these results predict a gain that is decreased by a factor of $n(\omega)$ compared to other non-collinear FWM simulations \cite{WeigandApplSci2015}. We do not account for the Raman response, which contributes to modulation instability \cite{BejotPRA2011}. Although we are in a regime for generating high harmonics \cite{KorobenkoOptExp2019}, we assume that the plasma-induced phase shift remains small compared to Kerr phase shift (eq. \ref{eq:pumpwave}). At higher intensities, it may limit the amplification, although optically induced damage can also occur \cite{XuOptComm2006}.

We solve eq. (\ref{eq:FME}) in the spatial-frequency domain, where the dispersion (the second term on the right hand side) and the nonlinearity (the third term) are integrated by a fourth-order Runge-Kutta (RK4) algorithm. We then Fourier transform the field to the momentum-frequency domain, making the approximation that the Fourier transform $\frac{\partial^2}{\partial x^2}$ in the spatial domain is $k_x^2$ in the momentum domain \cite{AgrawalNFO}. Unless otherwise stated, the nonlinear gain medium is 0.5~mm thick MgO and the pump parameters are: pulse duration $\Delta t = 800$~fs (Gaussian envelope full width at half-maximum, FWHM), peak intensity $I_p = 1.5\times10^{13}$~W/cm$^2$, beam waist $w_p = 400$~\textmu m (1/e$^2$ value), centered at $\lambda_p = 800$~nm.

In Fig. 1(b), we propagate our pulse to show the similarity in the gain curve in the KIA solution and our simulation, where the gain per length is $g = \ln\Bigl(S_f/S_i \Bigr)/L$ where $S_{i,f}$ are the peak of the initial and final spectra, respectively, and $L$ is the crystal thickness. To replicate the KIA results, the seed pulse in positive momentum-frequency space is uniform. Along the pump propagation axis near zero degrees, there is a saturation in the spectrum located at the pump wavelength. This saturation, caused by the pump beam size and bandwidth, is the main discrepancy observed between the KIA solution and the simulation. We find that the gain curve matches the KIA solution well, although we do not observe noticeable gain near the pump second harmonic, in contrast to a small amount predicted by KIA. We also note that pulse propagation slightly modifies the gain curve, where smaller pump $w_x$ leads to increased self-focusing, which in turn increases the angle of maximum amplification. Larger pump beams with longer pulse durations and thinner crystals, minimizing the effects of self-focusing and self-phase modulation, lead to improved agreement between the monochromatic plane wave solution of KIA and the simulations.

\section{Broadband amplification}

The results presented in Fig. 1 show that many frequencies can be amplified from a single pump, tuneable by the pump-seed angle and is wavelength dependent. In Fig. 2, we show the results of simulating the amplification of seed pulses with a finite pump pulse. In (a), we again overlay the KIA angle dependence when amplifying 100~fs seed pulses ($w_s = 50$~\textmu m) with 200~fs duration pump. For the 600~nm and 1000~nm seed cases, there is minimal spatial chirp, while the strong angle dependence at 1400~nm leads to noticeable spatial chirp. This spatial chirp is pump duration dependent, where longer pump pulses minimize the spatial chirp.

\begin{figure}[h]
\includegraphics[width=1\columnwidth]{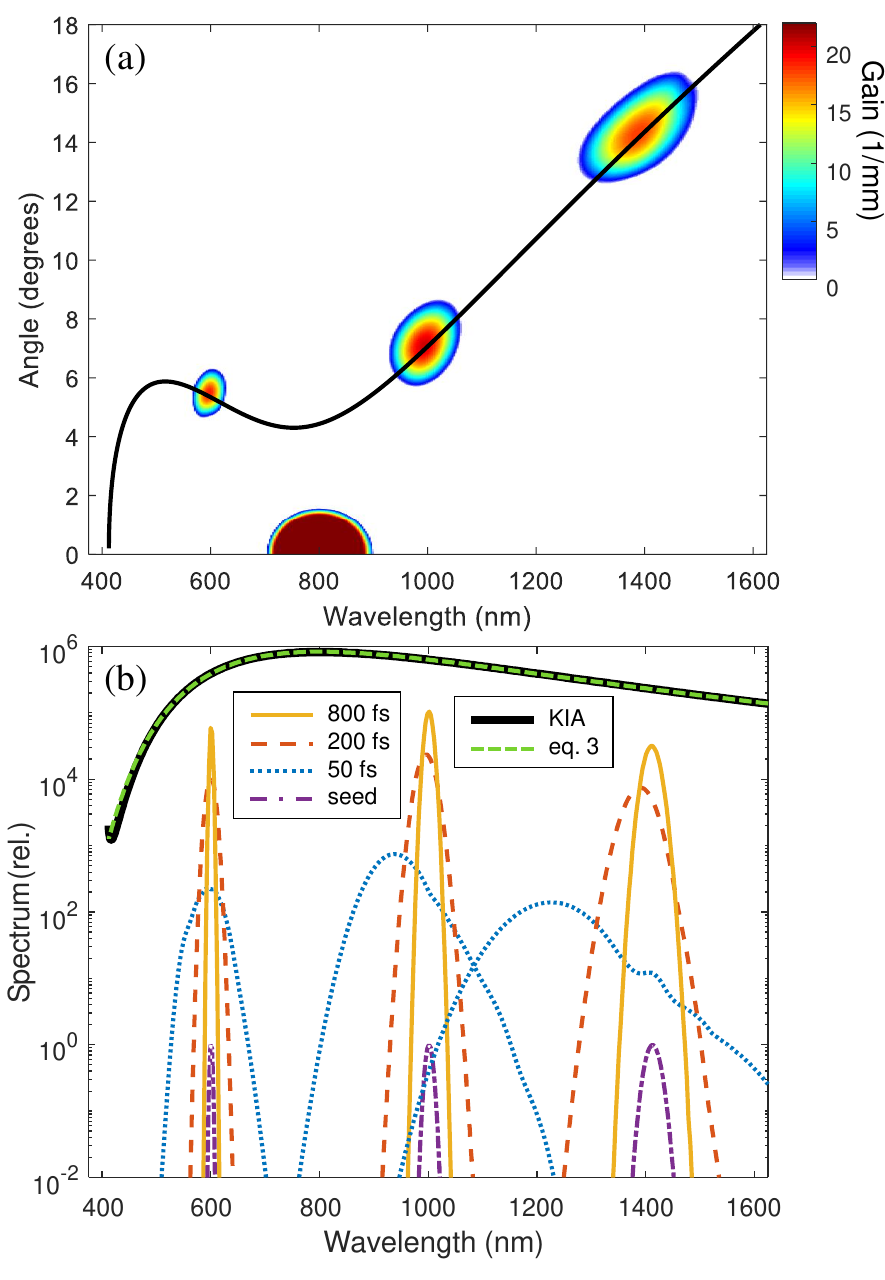}
\caption{(a) Angle dependence of amplification of 100~fs pulses with 200~fs pump follows the the curve predicted by KIA (solid black). The relatively short pump spatially chirps the amplified beam in the IR. (b) The amplified spectra of 600~nm (left), 1000~nm (middle), and 1400~nm (right) seed (dot-dashed purple). The 50~fs pump (dotted blue) is shorter in duration than the seed and causes significant spectral broadening and central wavelength shift. For the 200~fs case (long dashed orange), there is slight spectral broadening, and the central wavelength shift is more significant in the IR. Long 800~fs pump (solid yellow) amplification maintains the spectrum and beam profile for all seed wavelengths. Amplification predicted by KIA (solid black) compared to eq. 3 (short dashed green) agrees with the wavelength dependence.}
\end{figure}

We show the effect of the pump pulse duration on the amplified spectrum in (b). The seed spectrum of transform limited 100~fs pulses (dot-dashed purple) significantly broadens when the pump is shorter than the seed. This broadening can be understoond in the time domain because of the nonlinearity of the parametric amplification process. Because the amplification depends exponentially on the pump intensity, amplification is most efficient near the peak of the pump pulse. The result in the frequency domain is a broader pulse with decreased amplification, such as the 50~fs pump pulse cases (dotted blue). The amplified pulses are nearly transform-limited, leading to pulse compression. Pump pulses twice the seed duration (long dashed orange) show significantly more Gaussian shaped spectra, while 800~fs pump pulses (solid yellow) minimize any spectral distortion and limit spatial chirp. The longest pump pulse shows gain closest to the KIA predicted value ($g \approx$~23/mm compared to 26.7/mm at 1000~nm for KIA). Optimizing the gain is discussed below.

\begin{figure}[h]
\includegraphics[width=1\columnwidth]{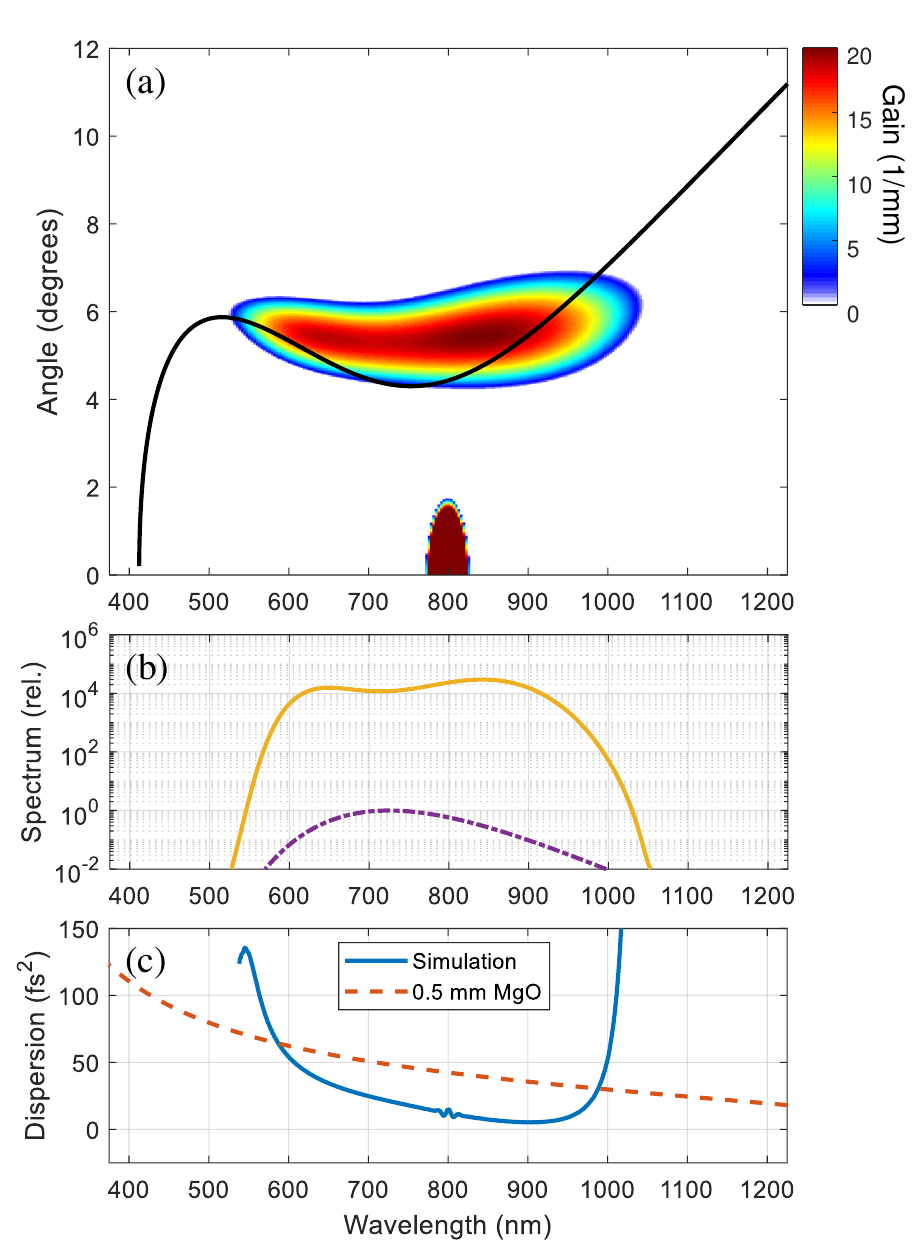}
\caption{Amplification of 5~fs pulse centred at 720~nm. (a) Angle dependence of amplification with seed at 5.4$^\circ$ leads to broadband amplification with negligible spatial chirp over the bandwidth (max gain from KIA in solid black). (b) Amplified broadband spectrum (solid yellow) compared to seed (dot-dashed purple) shows spectral shaping, but maintains the broad bandwidth. (c) GDD of the amplified pulse (solid blue) is less than the material dispersion (dashed orange) and is near zero for most of the bandwidth.}
\end{figure}

Because KIA predicts a near constant gain with little angle dependence about the pump wavelength, we simulate ultrashort pulse amplification in Fig. 3. The seed pulses are centered at 720~nm with 5~fs in duration (FWHM). In (a), because of the near-constant angle dependence about this spectral region, the amplified pulse is nearly spatially chirp-free. Again the KIA angle dependence is overlaid (solid black). In (b), we compare the seed spectrum (dot-dashed purple) with the amplified spectra (solid yellow); y axis is spectrum relative to the peak of the seed by taking the spectrum at $\theta_f = 5.4^\circ$. We calculate the resulting group delay dispersion (GDD) in (c), showing that the amplified seed accumulates less dispersion (solid blue) than propagation through the material (dashed orange) over most of its bandwidth.

\begin{figure}[h]
\includegraphics[width=1\columnwidth]{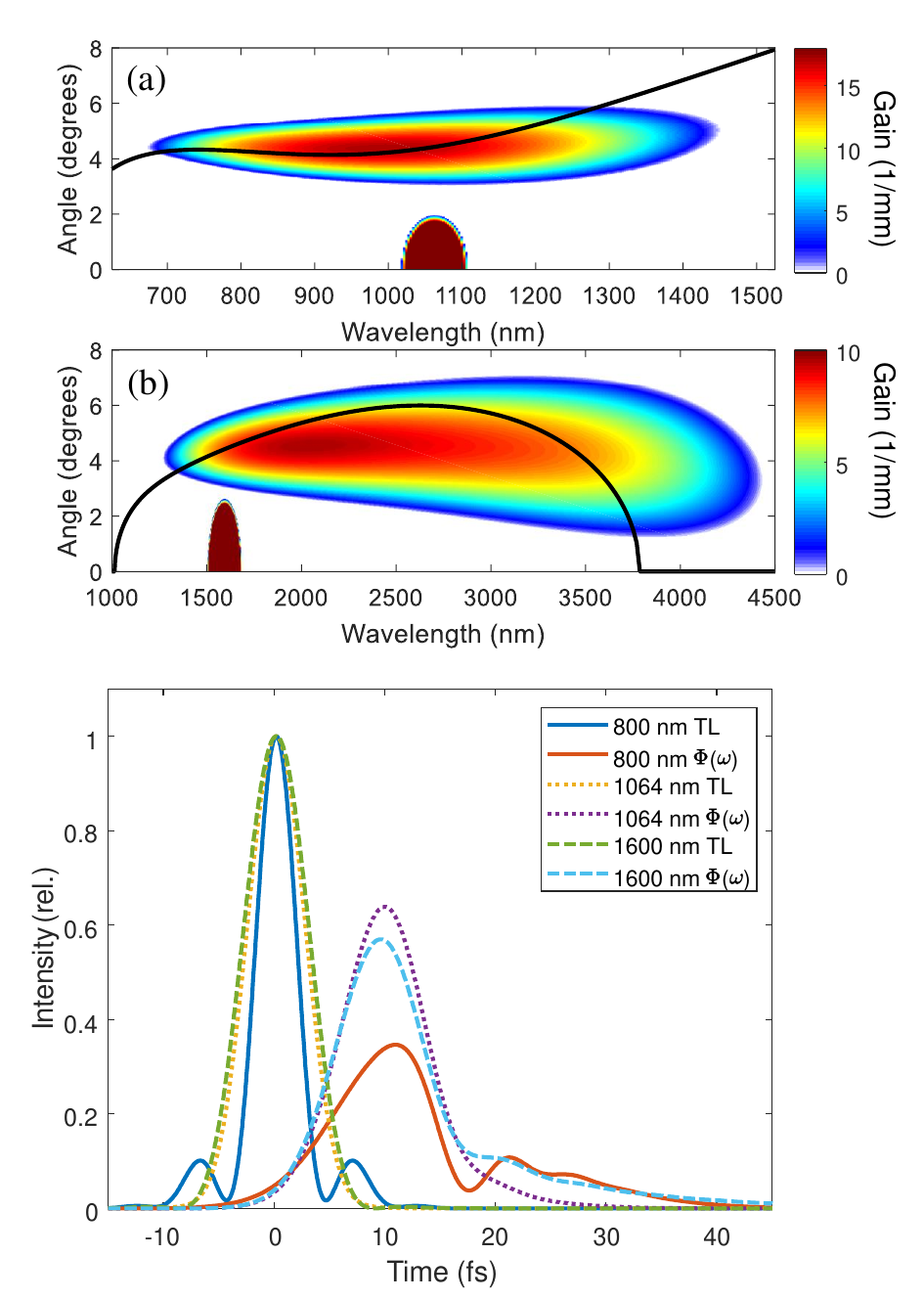}
\caption{Amplifying 5~fs pulses with (a) 1064~nm pump (970~nm seed), and (b) 1600~nm pump (2400~nm seed) (KIA angle dependence in solid black). (c) Because the spectrum is maintained in amplification, the transform limited (TL, peaks near 0~fs) pulse durations are comparable to the seed, becoming 4.1~fs for 800~nm pump (solid blue), 6.1~fs for 1064~nm pump (dotted yellow), and 6.7~fs for 1600~nm pump (dashed green). Although the dispersion is less than the material, the residual dispersion (including phase $\Phi(\omega)$, peaks near 10~fs) increases the pulse duration, becoming 10.1~fs for 800~nm (solid orange), 8.9~fs for 1064~nm (dotted purple), and 9.3~fs for 1600~nm (dashed blue).}
\end{figure}

We demonstrate that these broadband amplified pulses can be extended to the mid-infrared in Fig. 4. We simulate amplifying 5~fs seed pulses with (a) $\lambda_p =$~1064~nm (seed at $4.3^\circ$), and (b) $\lambda_p =$~1600~nm pump (seed at $4.5^\circ$); otherwise the pump parameters are the same as given previously. Both cases also demonstrate a supercontinuum amplification, although with diminished gain as predicted by eq. \ref{eq:gainest}. Although the angle dependence for the 1064~nm pump is similar to the 800~nm pump case, the 1600~nm pump case shows a significantly different angle dependence. In fact, without taking the pump intensity into account, FWM does not predict efficient amplification in this spectral region. Although the peak gain is near the pump wavelength, the amplified spectrum extends to beyond 4000~nm. 

We show the resulting amplified pulses in the time domain in (c). The transform limited (TL) pulse duration remains nearly unchanged after amplification, highlighting that gain narrowing is not significant. Because of the beneficial spectral shaping in the 800~nm pump case, the TL pulse duration decreases to 4.1~fs; however the longer pump wavelength cases increase to 6.1~fs and 6.7~fs for the 1064~nm and 1600~nm pump cases, respectively. Including the phase imparted by the amplification process, $\Phi(\omega)$, slightly increases the pulse duration, where the effect is most significant for the 800~nm pump case. Because the amplified pulse accumulates less GDD than it would by propagating through the material alone, the pulse duration only approximately doubles. If instead the material itself dominated the dispersion, then the pulse duration would be greater than 25~fs.

Moving forward, this amplification technique may be comparable to multi-pass amplification systems \cite{BackusRSI1998, BackusOptExp2017}, with inherent excellent pulse contrast \cite{KobayashiJPB2012, KobayashiIEEE2012, WangOL2019} and reduced complexity by avoiding active pulse selection for amplification. Furthermore, a mid-IR supercontinuum generated by $\omega-2\omega$ FWM in air is known to be carrier-envelope stable \cite{FujiOL2007, ChengOL2012, FujiApplSci2017}, but it has poor beam quality and is limited in power \cite{VoroninAPB2014}. Amplifying these mid-IR pulses may be useful as a source for condensed matter attosecond experiments \cite{ShiraiOL2018}.

\section{Amplification optimization}

The gain in the simulations is slightly lower than predicted by KIA, which did not take the creation of the idler into account. We calculate the gain in our simulations from the ratio of the amplified intensity to the seed, that is $g = \frac{1}{L} \ln \Big(\frac{I(L)}{I(0)} \Big)$, where $I(L) = I(\omega, k_{\perp}, L)$ and $I(0) = I(\omega_s, k_s, 0)$. In Fig. 5, we show the effect of propagation on a 100~fs seed pulse (the signal) focused to $w_s = 50$~\textmu m with a central wavelength of 1000~nm ($\lambda_p = 800$~nm). At $z=0$, the seed intensity is 6 orders of magnitude lower than the pump. In the first 100~\textmu m, there is little amplification of the signal because during this initial interaction, the pump and signal mix to create the idler. 

The signal and idler intensities are (in the undepleted pump approximation), respectively \cite{AgrawalNFO}
\begin{align}
I_s(z) &\approx I_s(0) \cosh^2 \Big(\frac{g}{2} z\Big) \\
I_i(z) &\approx \frac{\omega_i}{\omega_s} I_s(0) \sinh^2\Big(\frac{g}{2} z\Big).
\end{align}
Because the amplification is assumed to be exponential, which occurs when $\exp(- g z) \ll 1$, this initial interaction with the crystal decreases the calculated gain. We suggest that to more accurately predict the gain, assuming that $\exp(-g z) = 0.1$ is sufficiently small, then the initial $z = -\frac{1}{g}\ln(0.1) \approx 89$~\textmu m should be subtracted from $L$. This correction decreases the amplification by an order of magnitude. Once the signal and idler are approximately the same intensity, they both grow with a gain $g =$~25.8/mm (dashed black), near the predicted value given by KIA ($g =$~ 26.7/mm).

\begin{figure}[h]
\includegraphics[width=1\columnwidth]{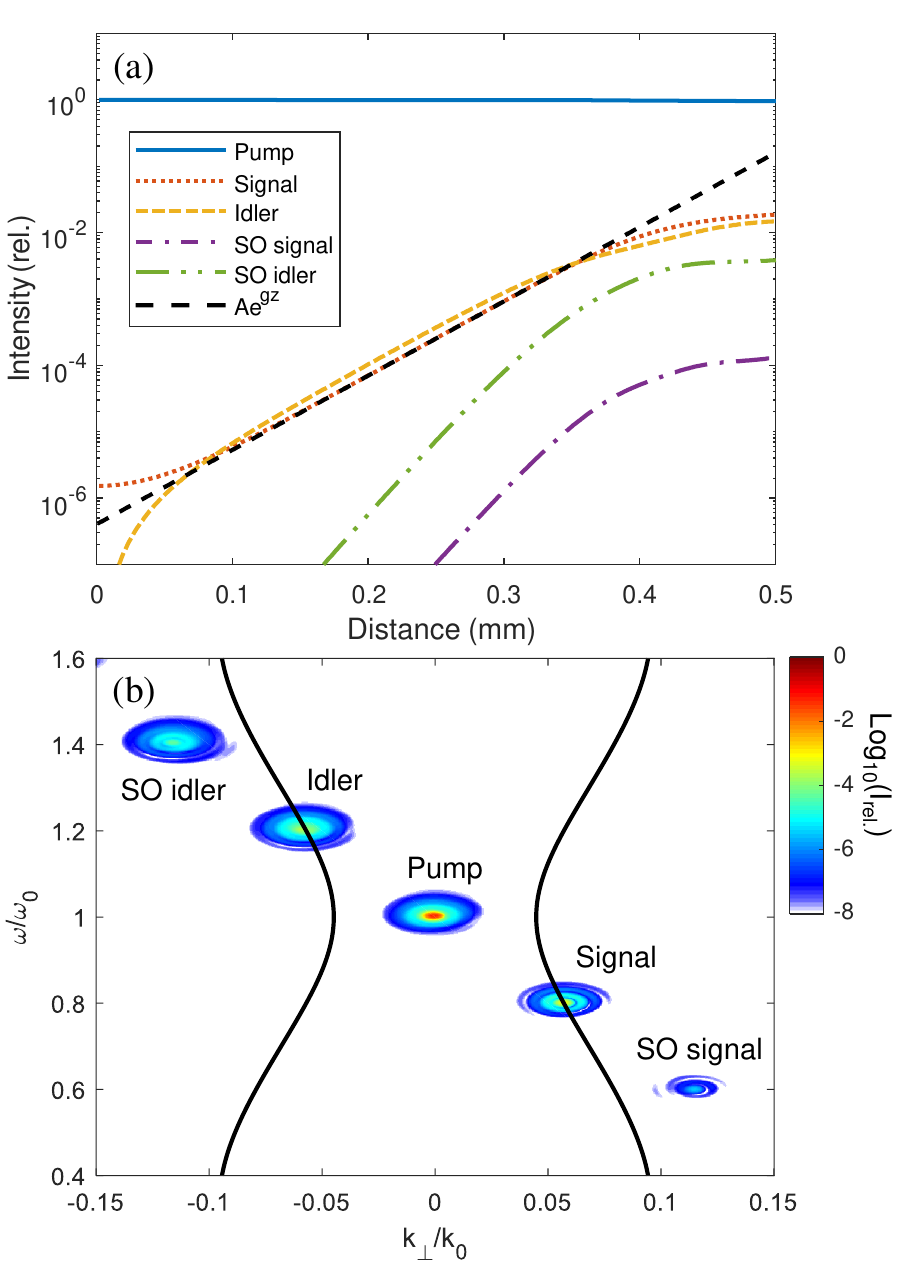}
\caption{The amplification of a 100~fs pulse through 0.5 mm MgO. (a) There is little amplification of the signal (dotted orange) within the first 100~\textmu m as the idler (dashed yellow) is created. Between 0.1 to 0.35~mm, $g = 25.8$/mm (dashed black). Amplification of the signal and idler saturates when they reach $\sim 1\%$ of the pump (solid blue) intensity; at this point higher order beamlets (second order SO signal, dot-dashed purple, and SO idler, double dot-dashed green) are of comparable intensity. (b) Amplified beams shown in $(k_\perp, \omega)$ space with KIA prediction (solid black).}
\end{figure}

Once the signal and idler are $\sim 1\%$ of the pump, they saturate and the second order (SO) signal and idler also saturate. Because of the pump depletion, the pump intensity decreases by $\sim 4\%$. These second order beamlets are from cascaded FWM with twice the gain. Further beamleats are then created, which can extend to the XUV \cite{HeApplSci2014, WeigandApplSci2015}.

When we reach saturation, these higher order beamlets are of comparable intensity to the signal and idler, as shown in Fig. 5(b). The signal, at $\omega/\omega_0 = 0.8$, has a transverse momentum that matches the maximum KIA gain curve (overlaid in solid black). We note that due to the symmetry of the gain, the idler is also located on the curve of maximum gain (phase matching) at $\omega/\omega_0 = 1.2$. The SO beamlets are located at symmetric points in $(k, \omega)$ space. Although we cannot directly amplify the second harmonic with KIA, it is possible to use the cascading FWM to create the second harmonic.

\begin{figure}[h]
\includegraphics[width=1\columnwidth]{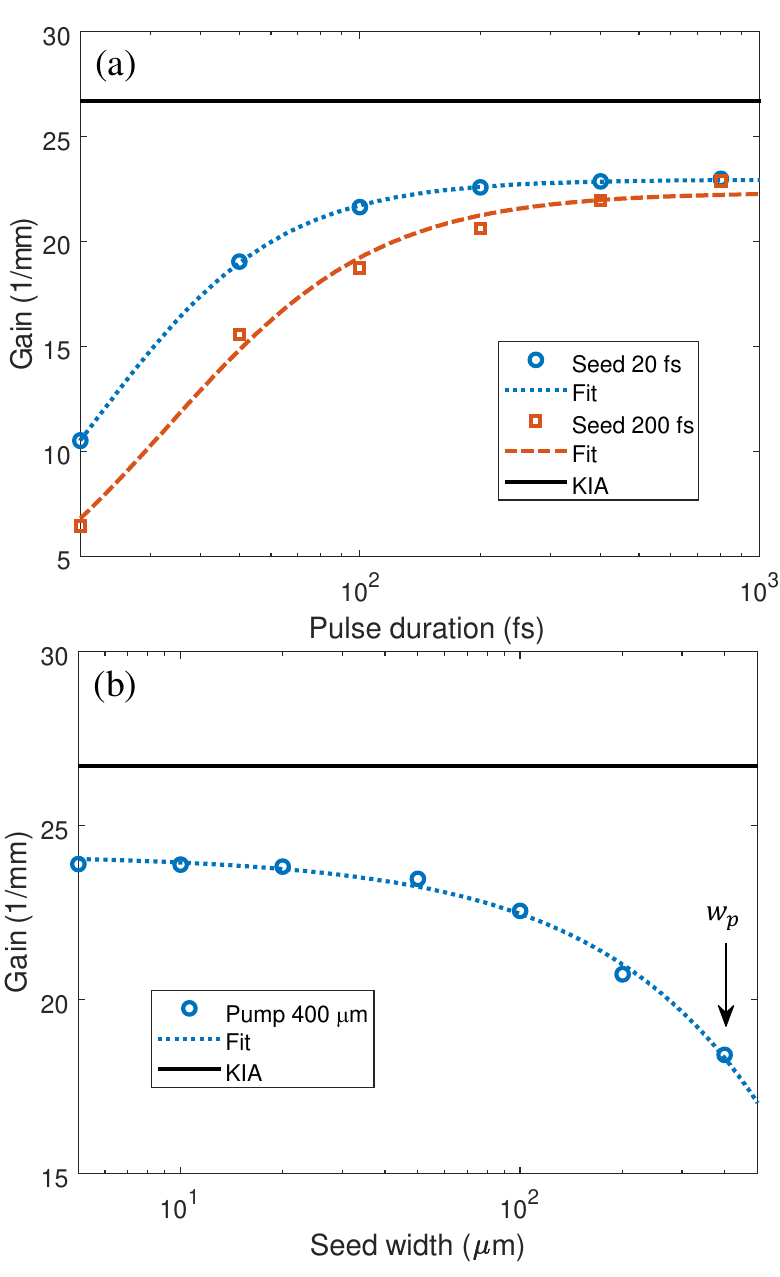}
\caption{Spatial and temporal optimization of the gain for finite pulses compared to KIA (solid black); self-phase modulation and self-focusing distort short, tightly focused pump pulses, limiting the gain. (a) Seed pulse duration of 20~fs (dotted blue) is more efficiently amplified by shorter pump pulses than seed of 200~fs in duration (dashed orange). (b) The seed waist beam also must be well within the pump beam waist, $w_p$.}
\end{figure}

As previously discussed, the amplification process is optimized when the seed pulse is well within the pump envelope, both temporally and spatially. In Fig. 6 we discuss optimizing the gain by varying the pump pulse duration and the seed beam width. In (a), we investigate the gain as a function of pump pulse duration for 20~fs (dotted blue) and 200~fs (dashed orange) seed pulses. To calculate the gain, we integrate over the amplified spectrum and fit the curve to 
\begin{align}
\frac{1}{f(\tau_p)} = \frac{1}{a \tau^b} + \frac{1}{c},
\end{align}
where $c =$ 23.0/mm for $\tau_s = 20$~fs and $c =$ 22.3/mm for $\tau_s =$ 200~fs. The gain predicted by KIA ($g\approx 26.7$~/mm) is shown for reference. The gain depends on the pump pulse duration $\tau_{p}$ when $\tau_{p} \lesssim 5\times \tau_{s}$, and then approaches a maximum limit. 

We also investigate the spatial overlap to optimize the amplification in (b), with the pump waist $w_p = 400$~\textmu m and pump pulse duration $\Delta t =$ 800~fs. We fit this gain as a function of seed width to the function
\begin{align}
f(w_s) = a - \bigg(\frac{w_s}{b} \bigg)^c,
\end{align}
where $a =$ 24.2/mm and $b =$ 55.5~\textmu m. When the seed waist, $w_s \lesssim 5 w_p$, the seed is most efficiently amplified. We note that for longer crystals, amplification is less sensitive to the seed parameter because the portion of the seed that is spatially and temporally within the peak of the pump envelope is efficiently amplified. Although propagating in 2D decreases the spatial dispersion of the beam, effectively increasing the Rayleigh range, we find that this effect only plays a significant role for seed beams smaller than 30~\textmu m. In 3D, the larger seed beam divergence may lead to a decrease in amplification for beams of such small beam waists.

\section{Conclusions}
In conclusion, we compare the amplification predicted by Kerr instability with a 2D propagation simulation. We find good agreement between previously-developed KIA theory and our simulation, where disagreement stems largely from the finite pump beam size and pulse duration, as well as pump pulse propagation effects. We find agreement for the angle dependence between KIA and FWM when the nonlinear Kerr coefficient is considered. We also find a frequency dependence of the gain, leading the way for modified parametric amplification equations. The amplification of long pulses leads to substantial spectral broadening and pulse compression. We find it is possible to amplify few-cycle pulses without spatial chirp when the seed wavelength is near the pump. We also note that this broadband amplification is not limited to the near infrared, but is also achievable with mid-IR pump sources and can amplify single-cycle pulses.

\begin{acknowledgments}
SG acknowledges funding from Mitacs Globalink Research Internship. We acknowledge funding from Natural Sciences and Engineering Research Council of Canada (RGPIN-2019-06877) and the University of Windsor Xcellerate grant (5218522). TJH thanks Giulio Vampa, Thomas Brabec, and Claire Duncan for useful conversations.
\end{acknowledgments}

\appendix
\section{Kerr Instability Amplification Equations}

We use the equations from Nesrallah \textit{et al} \cite{NesrallahOptica2018} to generate the Kerr instability amplification gain curve of Fig. 1(a). We summarize and transcribe the necessary equations for reproducing these curves here. 

The wavelength and angular dependence of maximum gain (solid black line) is given by,
\begin{equation}
g_{max} = \frac{k_p \delta_\perp^2}{k_p^2 - \sigma^2 D_u^2}.
\end{equation}
The transverse instability half-width is,
\begin{equation}
\delta_\perp^2 = \frac{k_{n-} k_{n+} \sqrt{k_p^2 - \sigma^2 D_u^2}}{k_p},
\end{equation}
where,
\begin{equation}
\sigma = \frac{k_v + D_g}{k_p}.
\end{equation}
The nonlinear wavevector is
\begin{equation}
k_{n\pm} = \frac{\omega_p \pm \Omega}{c}\sqrt{n_2 I_p},
\end{equation}
where,
\begin{equation}
\Omega = \omega - \omega_p.
\end{equation}
The perturbation wavevector is given by,
\begin{equation}
k_v^2 = \frac{\omega^2}{c^2}(n^2 + 2 n_2 I_p).
\end{equation}
The odd and even dispersion relations are,
\begin{align}
D_u &= \frac{1}{c}(\eta_p \Omega + \eta_g \Omega + \eta_u \omega_p) \\
D_g &= \frac{1}{c}(\eta_g \omega_p + \eta_u \Omega),
\end{align}
respectively, allowing for dispersion to be accounted for without approximations. The refractive index, in this notation, becomes
\begin{equation}
\eta = \sqrt{n^2 + 2 n_2 I_p},
\end{equation}
which has also been separated into even and odd components without approximation,
\begin{equation}
\eta_{g,u} = \frac{1}{2} \Bigl( \Delta \eta(\Omega) \pm \Delta \eta(-\Omega) \Bigr),
\end{equation}
where
\begin{equation}
\Delta \eta = \eta(\omega_p + \Omega) - \eta_p.
\end{equation}

The gain amplitude is,
\begin{equation}
g = -2 \Im\{K_g\},
\end{equation}
where $\Im$ is the imaginary part of
\begin{equation}
K_g = -\frac{1}{2} \frac{k_p \sqrt{(\kappa_\perp^2 - k_\perp^2)^2 - \delta_\perp^4}}{k_p^2 - \sigma^2 D_u^2}
\end{equation}
and
\begin{equation}
\kappa_\perp^2 = (k_p^2 - D_u^2)(\sigma^2 - 1).
\end{equation}
The seed transverse momentum, $k_\perp$, is the array of possible seed transverse momentum values.


\begin{thebibliography}{47}%
\makeatletter
\providecommand \@ifxundefined [1]{%
 \@ifx{#1\undefined}
}%
\providecommand \@ifnum [1]{%
 \ifnum #1\expandafter \@firstoftwo
 \else \expandafter \@secondoftwo
 \fi
}%
\providecommand \@ifx [1]{%
 \ifx #1\expandafter \@firstoftwo
 \else \expandafter \@secondoftwo
 \fi
}%
\providecommand \natexlab [1]{#1}%
\providecommand \enquote  [1]{``#1''}%
\providecommand \bibnamefont  [1]{#1}%
\providecommand \bibfnamefont [1]{#1}%
\providecommand \citenamefont [1]{#1}%
\providecommand \href@noop [0]{\@secondoftwo}%
\providecommand \href [0]{\begingroup \@sanitize@url \@href}%
\providecommand \@href[1]{\@@startlink{#1}\@@href}%
\providecommand \@@href[1]{\endgroup#1\@@endlink}%
\providecommand \@sanitize@url [0]{\catcode `\\12\catcode `\$12\catcode
  `\&12\catcode `\#12\catcode `\^12\catcode `\_12\catcode `\%12\relax}%
\providecommand \@@startlink[1]{}%
\providecommand \@@endlink[0]{}%
\providecommand \url  [0]{\begingroup\@sanitize@url \@url }%
\providecommand \@url [1]{\endgroup\@href {#1}{\urlprefix }}%
\providecommand \urlprefix  [0]{URL }%
\providecommand \Eprint [0]{\href }%
\providecommand \doibase [0]{https://doi.org/}%
\providecommand \selectlanguage [0]{\@gobble}%
\providecommand \bibinfo  [0]{\@secondoftwo}%
\providecommand \bibfield  [0]{\@secondoftwo}%
\providecommand \translation [1]{[#1]}%
\providecommand \BibitemOpen [0]{}%
\providecommand \bibitemStop [0]{}%
\providecommand \bibitemNoStop [0]{.\EOS\space}%
\providecommand \EOS [0]{\spacefactor3000\relax}%
\providecommand \BibitemShut  [1]{\csname bibitem#1\endcsname}%
\let\auto@bib@innerbib\@empty
\bibitem [{\citenamefont {Calegari}\ \emph {et~al.}(2016)\citenamefont
  {Calegari}, \citenamefont {Sansone}, \citenamefont {Stagira}, \citenamefont
  {Vozzi},\ and\ \citenamefont {Nisoli}}]{CalegariJPB2016}%
  \BibitemOpen
  \bibfield  {author} {\bibinfo {author} {\bibfnamefont {F.}~\bibnamefont
  {Calegari}}, \bibinfo {author} {\bibfnamefont {G.}~\bibnamefont {Sansone}},
  \bibinfo {author} {\bibfnamefont {S.}~\bibnamefont {Stagira}}, \bibinfo
  {author} {\bibfnamefont {C.}~\bibnamefont {Vozzi}},\ and\ \bibinfo {author}
  {\bibfnamefont {M.}~\bibnamefont {Nisoli}},\ }\bibfield  {title} {\bibinfo
  {title} {Advances in attosecond science},\ }\href@noop {} {\bibfield
  {journal} {\bibinfo  {journal} {J. Phys. B: At. Mol. Opt. Phys.}\ }\textbf
  {\bibinfo {volume} {49}},\ \bibinfo {pages} {062001} (\bibinfo {year}
  {2016})}\BibitemShut {NoStop}%
\bibitem [{\citenamefont {Seres}\ \emph {et~al.}(2003)\citenamefont {Seres},
  \citenamefont {M\"{u}ller}, \citenamefont {Seres}, \citenamefont {O'Keeffe},
  \citenamefont {Lenner}, \citenamefont {Herzog}, \citenamefont {Kaplan},
  \citenamefont {Spielmann},\ and\ \citenamefont {Krausz}}]{SeresOL2003}%
  \BibitemOpen
  \bibfield  {author} {\bibinfo {author} {\bibfnamefont {J.}~\bibnamefont
  {Seres}}, \bibinfo {author} {\bibfnamefont {A.}~\bibnamefont {M\"{u}ller}},
  \bibinfo {author} {\bibfnamefont {E.}~\bibnamefont {Seres}}, \bibinfo
  {author} {\bibfnamefont {K.}~\bibnamefont {O'Keeffe}}, \bibinfo {author}
  {\bibfnamefont {M.}~\bibnamefont {Lenner}}, \bibinfo {author} {\bibfnamefont
  {R.}~\bibnamefont {Herzog}}, \bibinfo {author} {\bibfnamefont
  {D.}~\bibnamefont {Kaplan}}, \bibinfo {author} {\bibfnamefont
  {C.}~\bibnamefont {Spielmann}},\ and\ \bibinfo {author} {\bibfnamefont
  {F.}~\bibnamefont {Krausz}},\ }\bibfield  {title} {\bibinfo {title}
  {Sub-10-fs, terawatt-scale ti:sapphire laser system},\ }\href@noop {}
  {\bibfield  {journal} {\bibinfo  {journal} {Opt. Lett.}\ }\textbf {\bibinfo
  {volume} {28}},\ \bibinfo {pages} {1832} (\bibinfo {year}
  {2003})}\BibitemShut {NoStop}%
\bibitem [{\citenamefont {Weiner}(2011)}]{WeinerOC2011}%
  \BibitemOpen
  \bibfield  {author} {\bibinfo {author} {\bibfnamefont {A.}~\bibnamefont
  {Weiner}},\ }\bibfield  {title} {\bibinfo {title} {Ultrafast optical pulse
  shaping: A tutorial review},\ }\href@noop {} {\bibfield  {journal} {\bibinfo
  {journal} {Opt. Commun.}\ }\textbf {\bibinfo {volume} {284}},\ \bibinfo
  {pages} {3669} (\bibinfo {year} {2011})}\BibitemShut {NoStop}%
\bibitem [{\citenamefont {Liu}\ \emph {et~al.}(2020)\citenamefont {Liu},
  \citenamefont {Wang}, \citenamefont {Wang}, \citenamefont {Lu}, \citenamefont
  {Bai}, \citenamefont {Liu}, \citenamefont {Li}, \citenamefont {Liu},
  \citenamefont {Yu}, \citenamefont {Leng},\ and\ \citenamefont
  {Li}}]{LiuOptics2020}%
  \BibitemOpen
  \bibfield  {author} {\bibinfo {author} {\bibfnamefont {X.}~\bibnamefont
  {Liu}}, \bibinfo {author} {\bibfnamefont {C.}~\bibnamefont {Wang}}, \bibinfo
  {author} {\bibfnamefont {X.}~\bibnamefont {Wang}}, \bibinfo {author}
  {\bibfnamefont {X.}~\bibnamefont {Lu}}, \bibinfo {author} {\bibfnamefont
  {P.}~\bibnamefont {Bai}}, \bibinfo {author} {\bibfnamefont {Y.}~\bibnamefont
  {Liu}}, \bibinfo {author} {\bibfnamefont {Y.}~\bibnamefont {Li}}, \bibinfo
  {author} {\bibfnamefont {K.}~\bibnamefont {Liu}}, \bibinfo {author}
  {\bibfnamefont {L.}~\bibnamefont {Yu}}, \bibinfo {author} {\bibfnamefont
  {Y.}~\bibnamefont {Leng}},\ and\ \bibinfo {author} {\bibfnamefont
  {R.}~\bibnamefont {Li}},\ }\bibfield  {title} {\bibinfo {title} {Dispersion
  management in a 10-pw laser front end},\ }\href@noop {} {\bibfield  {journal}
  {\bibinfo  {journal} {Optics}\ }\textbf {\bibinfo {volume} {1}},\ \bibinfo
  {pages} {191} (\bibinfo {year} {2020})}\BibitemShut {NoStop}%
\bibitem [{\citenamefont {Kobayashi}\ \emph
  {et~al.}(2012{\natexlab{a}})\citenamefont {Kobayashi}, \citenamefont {Liu},\
  and\ \citenamefont {Okamura}}]{KobayashiJPB2012}%
  \BibitemOpen
  \bibfield  {author} {\bibinfo {author} {\bibfnamefont {T.}~\bibnamefont
  {Kobayashi}}, \bibinfo {author} {\bibfnamefont {J.}~\bibnamefont {Liu}},\
  and\ \bibinfo {author} {\bibfnamefont {K.}~\bibnamefont {Okamura}},\
  }\bibfield  {title} {\bibinfo {title} {Applications of parametric processes
  to high-quality multicolour ultrashort pulses, pulse cleaning and cep stable
  sub-3fs pulse},\ }\href@noop {} {\bibfield  {journal} {\bibinfo  {journal}
  {J. Phys. B: At. Mol. and Opt. Phys.}\ }\textbf {\bibinfo {volume} {45}},\
  \bibinfo {pages} {074005} (\bibinfo {year} {2012}{\natexlab{a}})}\BibitemShut
  {NoStop}%
\bibitem [{\citenamefont {Fattahi}\ \emph {et~al.}(2014)\citenamefont
  {Fattahi}, \citenamefont {Barros}, \citenamefont {Gorjan}, \citenamefont
  {Nubbemeyer}, \citenamefont {Alsaif}, \citenamefont {Teisset}, \citenamefont
  {Schultze}, \citenamefont {Prinz}, \citenamefont {Haefner}, \citenamefont
  {Ueffing}, \citenamefont {Alismail}, \citenamefont {V{\'a}mos}, \citenamefont
  {Schwarz}, \citenamefont {Pronin}, \citenamefont {Brons}, \citenamefont
  {Geng}, \citenamefont {Arisholm}, \citenamefont {Ciappina}, \citenamefont
  {Yakovlev}, \citenamefont {Kim}, \citenamefont {Azzeer}, \citenamefont
  {Karpowicz}, \citenamefont {Sutter}, \citenamefont {Major}, \citenamefont
  {Metzger},\ and\ \citenamefont {Krausz}}]{FattahiOptica2014}%
  \BibitemOpen
  \bibfield  {author} {\bibinfo {author} {\bibfnamefont {H.}~\bibnamefont
  {Fattahi}}, \bibinfo {author} {\bibfnamefont {H.~G.}\ \bibnamefont {Barros}},
  \bibinfo {author} {\bibfnamefont {M.}~\bibnamefont {Gorjan}}, \bibinfo
  {author} {\bibfnamefont {T.}~\bibnamefont {Nubbemeyer}}, \bibinfo {author}
  {\bibfnamefont {B.}~\bibnamefont {Alsaif}}, \bibinfo {author} {\bibfnamefont
  {C.~Y.}\ \bibnamefont {Teisset}}, \bibinfo {author} {\bibfnamefont
  {M.}~\bibnamefont {Schultze}}, \bibinfo {author} {\bibfnamefont
  {S.}~\bibnamefont {Prinz}}, \bibinfo {author} {\bibfnamefont
  {M.}~\bibnamefont {Haefner}}, \bibinfo {author} {\bibfnamefont
  {M.}~\bibnamefont {Ueffing}}, \bibinfo {author} {\bibfnamefont
  {A.}~\bibnamefont {Alismail}}, \bibinfo {author} {\bibfnamefont
  {L.}~\bibnamefont {V{\'a}mos}}, \bibinfo {author} {\bibfnamefont
  {A.}~\bibnamefont {Schwarz}}, \bibinfo {author} {\bibfnamefont
  {O.}~\bibnamefont {Pronin}}, \bibinfo {author} {\bibfnamefont
  {J.}~\bibnamefont {Brons}}, \bibinfo {author} {\bibfnamefont {X.~T.}\
  \bibnamefont {Geng}}, \bibinfo {author} {\bibfnamefont {G.}~\bibnamefont
  {Arisholm}}, \bibinfo {author} {\bibfnamefont {M.}~\bibnamefont {Ciappina}},
  \bibinfo {author} {\bibfnamefont {V.~S.}\ \bibnamefont {Yakovlev}}, \bibinfo
  {author} {\bibfnamefont {D.~E.}\ \bibnamefont {Kim}}, \bibinfo {author}
  {\bibfnamefont {A.~M.}\ \bibnamefont {Azzeer}}, \bibinfo {author}
  {\bibfnamefont {N.}~\bibnamefont {Karpowicz}}, \bibinfo {author}
  {\bibfnamefont {D.}~\bibnamefont {Sutter}}, \bibinfo {author} {\bibfnamefont
  {Z.}~\bibnamefont {Major}}, \bibinfo {author} {\bibfnamefont
  {T.}~\bibnamefont {Metzger}},\ and\ \bibinfo {author} {\bibfnamefont
  {F.}~\bibnamefont {Krausz}},\ }\bibfield  {title} {\bibinfo {title}
  {Third-generation femtosecond technology},\ }\href@noop {} {\bibfield
  {journal} {\bibinfo  {journal} {Optica}\ }\textbf {\bibinfo {volume} {1}},\
  \bibinfo {pages} {45 } (\bibinfo {year} {2014})}\BibitemShut {NoStop}%
\bibitem [{\citenamefont {Popmintchev}\ \emph {et~al.}(2012)\citenamefont
  {Popmintchev}, \citenamefont {Chen}, \citenamefont {Popmintchev},
  \citenamefont {Arpin}, \citenamefont {Brown}, \citenamefont {Ali\v{s}auskas},
  \citenamefont {Andriukaitis}, \citenamefont {Bal\v{c}iunas}, \citenamefont
  {M\"{u}cke}, \citenamefont {Pugzlys}, \citenamefont {Baltu\v{s}ka},
  \citenamefont {Shim}, \citenamefont {Schrauth}, \citenamefont {Gaeta},
  \citenamefont {Hern\'{a}ndez-Garc\'{i}a}, \citenamefont {Plaja},
  \citenamefont {Becker}, \citenamefont {Jaron-Becker}, \citenamefont
  {Murnane},\ and\ \citenamefont {Kapteyn}}]{PopmintchevScience2012}%
  \BibitemOpen
  \bibfield  {author} {\bibinfo {author} {\bibfnamefont {T.}~\bibnamefont
  {Popmintchev}}, \bibinfo {author} {\bibfnamefont {M.}~\bibnamefont {Chen}},
  \bibinfo {author} {\bibfnamefont {D.}~\bibnamefont {Popmintchev}}, \bibinfo
  {author} {\bibfnamefont {P.}~\bibnamefont {Arpin}}, \bibinfo {author}
  {\bibfnamefont {S.}~\bibnamefont {Brown}}, \bibinfo {author} {\bibfnamefont
  {S.}~\bibnamefont {Ali\v{s}auskas}}, \bibinfo {author} {\bibfnamefont
  {G.}~\bibnamefont {Andriukaitis}}, \bibinfo {author} {\bibfnamefont
  {T.}~\bibnamefont {Bal\v{c}iunas}}, \bibinfo {author} {\bibfnamefont
  {O.}~\bibnamefont {M\"{u}cke}}, \bibinfo {author} {\bibfnamefont
  {A.}~\bibnamefont {Pugzlys}}, \bibinfo {author} {\bibfnamefont
  {A.}~\bibnamefont {Baltu\v{s}ka}}, \bibinfo {author} {\bibfnamefont
  {B.}~\bibnamefont {Shim}}, \bibinfo {author} {\bibfnamefont {S.}~\bibnamefont
  {Schrauth}}, \bibinfo {author} {\bibfnamefont {A.}~\bibnamefont {Gaeta}},
  \bibinfo {author} {\bibfnamefont {C.}~\bibnamefont
  {Hern\'{a}ndez-Garc\'{i}a}}, \bibinfo {author} {\bibfnamefont
  {L.}~\bibnamefont {Plaja}}, \bibinfo {author} {\bibfnamefont
  {A.}~\bibnamefont {Becker}}, \bibinfo {author} {\bibfnamefont
  {A.}~\bibnamefont {Jaron-Becker}}, \bibinfo {author} {\bibfnamefont
  {M.}~\bibnamefont {Murnane}},\ and\ \bibinfo {author} {\bibfnamefont
  {H.}~\bibnamefont {Kapteyn}},\ }\bibfield  {title} {\bibinfo {title} {Bright
  coherent ultrahigh harmonics in the kev x-ray regime from mid-infrared
  femtosecond lasers},\ }\href@noop {} {\bibfield  {journal} {\bibinfo
  {journal} {Science}\ }\textbf {\bibinfo {volume} {336}},\ \bibinfo {pages}
  {1287} (\bibinfo {year} {2012})}\BibitemShut {NoStop}%
\bibitem [{\citenamefont {Ghimire}\ \emph {et~al.}(2011)\citenamefont
  {Ghimire}, \citenamefont {DiChiara}, \citenamefont {Sistrunk}, \citenamefont
  {Agostini}, \citenamefont {DiMauro},\ and\ \citenamefont
  {Reis}}]{GhimireNP2011}%
  \BibitemOpen
  \bibfield  {author} {\bibinfo {author} {\bibfnamefont {S.}~\bibnamefont
  {Ghimire}}, \bibinfo {author} {\bibfnamefont {A.}~\bibnamefont {DiChiara}},
  \bibinfo {author} {\bibfnamefont {E.}~\bibnamefont {Sistrunk}}, \bibinfo
  {author} {\bibfnamefont {P.}~\bibnamefont {Agostini}}, \bibinfo {author}
  {\bibfnamefont {L.}~\bibnamefont {DiMauro}},\ and\ \bibinfo {author}
  {\bibfnamefont {D.}~\bibnamefont {Reis}},\ }\bibfield  {title} {\bibinfo
  {title} {Observation of high-order harmonic generation in a bulk crystal},\
  }\href@noop {} {\bibfield  {journal} {\bibinfo  {journal} {Nat. Phys.}\
  }\textbf {\bibinfo {volume} {7}},\ \bibinfo {pages} {138} (\bibinfo {year}
  {2011})}\BibitemShut {NoStop}%
\bibitem [{\citenamefont {Vampa}\ \emph {et~al.}(2015)\citenamefont {Vampa},
  \citenamefont {Hammond}, \citenamefont {Thir\'{e}}, \citenamefont {Schmidt},
  \citenamefont {L\'{e}gar\'{e}}, \citenamefont {McDonald}, \citenamefont
  {Brabec},\ and\ \citenamefont {Corkum}}]{VampaNature2015}%
  \BibitemOpen
  \bibfield  {author} {\bibinfo {author} {\bibfnamefont {G.}~\bibnamefont
  {Vampa}}, \bibinfo {author} {\bibfnamefont {T.}~\bibnamefont {Hammond}},
  \bibinfo {author} {\bibfnamefont {N.}~\bibnamefont {Thir\'{e}}}, \bibinfo
  {author} {\bibfnamefont {B.}~\bibnamefont {Schmidt}}, \bibinfo {author}
  {\bibfnamefont {F.}~\bibnamefont {L\'{e}gar\'{e}}}, \bibinfo {author}
  {\bibfnamefont {C.}~\bibnamefont {McDonald}}, \bibinfo {author}
  {\bibfnamefont {T.}~\bibnamefont {Brabec}},\ and\ \bibinfo {author}
  {\bibfnamefont {P.}~\bibnamefont {Corkum}},\ }\bibfield  {title} {\bibinfo
  {title} {Linking high harmonics from gases and solids},\ }\href@noop {}
  {\bibfield  {journal} {\bibinfo  {journal} {Nature}\ }\textbf {\bibinfo
  {volume} {522}},\ \bibinfo {pages} {462} (\bibinfo {year}
  {2015})}\BibitemShut {NoStop}%
\bibitem [{\citenamefont {Gallmann}\ \emph {et~al.}(2012)\citenamefont
  {Gallmann}, \citenamefont {Cirelli},\ and\ \citenamefont
  {Keller}}]{GallmannARP2012}%
  \BibitemOpen
  \bibfield  {author} {\bibinfo {author} {\bibfnamefont {L.}~\bibnamefont
  {Gallmann}}, \bibinfo {author} {\bibfnamefont {C.}~\bibnamefont {Cirelli}},\
  and\ \bibinfo {author} {\bibfnamefont {U.}~\bibnamefont {Keller}},\
  }\bibfield  {title} {\bibinfo {title} {Attosecond science: Recent highlights
  and future trends},\ }\href@noop {} {\bibfield  {journal} {\bibinfo
  {journal} {Annu. Rev. Phys. Chem.}\ }\textbf {\bibinfo {volume} {63}},\
  \bibinfo {pages} {447} (\bibinfo {year} {2012})}\BibitemShut {NoStop}%
\bibitem [{\citenamefont {Nisoli}\ \emph {et~al.}(2017)\citenamefont {Nisoli},
  \citenamefont {Decleva}, \citenamefont {Calegari}, \citenamefont {Palacios},\
  and\ \citenamefont {Mart\'{i}n}}]{NisoliCR2017}%
  \BibitemOpen
  \bibfield  {author} {\bibinfo {author} {\bibfnamefont {M.}~\bibnamefont
  {Nisoli}}, \bibinfo {author} {\bibfnamefont {P.}~\bibnamefont {Decleva}},
  \bibinfo {author} {\bibfnamefont {F.}~\bibnamefont {Calegari}}, \bibinfo
  {author} {\bibfnamefont {A.}~\bibnamefont {Palacios}},\ and\ \bibinfo
  {author} {\bibfnamefont {F.}~\bibnamefont {Mart\'{i}n}},\ }\bibfield  {title}
  {\bibinfo {title} {Attosecond electron dynamics in molecules},\ }\href@noop
  {} {\bibfield  {journal} {\bibinfo  {journal} {Chem. Rev.}\ }\textbf
  {\bibinfo {volume} {117}},\ \bibinfo {pages} {10760} (\bibinfo {year}
  {2017})}\BibitemShut {NoStop}%
\bibitem [{\citenamefont {Li}\ \emph {et~al.}(2020)\citenamefont {Li},
  \citenamefont {Lu}, \citenamefont {Chew}, \citenamefont {Han}, \citenamefont
  {Li}, \citenamefont {Wu}, \citenamefont {Wang}, \citenamefont {Ghimire},\
  and\ \citenamefont {Chang}}]{LiNC2020}%
  \BibitemOpen
  \bibfield  {author} {\bibinfo {author} {\bibfnamefont {J.}~\bibnamefont
  {Li}}, \bibinfo {author} {\bibfnamefont {J.}~\bibnamefont {Lu}}, \bibinfo
  {author} {\bibfnamefont {A.}~\bibnamefont {Chew}}, \bibinfo {author}
  {\bibfnamefont {S.}~\bibnamefont {Han}}, \bibinfo {author} {\bibfnamefont
  {J.}~\bibnamefont {Li}}, \bibinfo {author} {\bibfnamefont {Y.}~\bibnamefont
  {Wu}}, \bibinfo {author} {\bibfnamefont {H.}~\bibnamefont {Wang}}, \bibinfo
  {author} {\bibfnamefont {S.}~\bibnamefont {Ghimire}},\ and\ \bibinfo {author}
  {\bibfnamefont {Z.}~\bibnamefont {Chang}},\ }\bibfield  {title} {\bibinfo
  {title} {Attosecond science based on high harmonic generation from gases to
  solids},\ }\href@noop {} {\bibfield  {journal} {\bibinfo  {journal} {Nat.
  Commun.}\ }\textbf {\bibinfo {volume} {11}},\ \bibinfo {pages} {2748}
  (\bibinfo {year} {2020})}\BibitemShut {NoStop}%
\bibitem [{\citenamefont {Nisoli}\ \emph {et~al.}(1997)\citenamefont {Nisoli},
  \citenamefont {Silvestri}, \citenamefont {Svelto}, \citenamefont
  {Szip\"{o}cs}, \citenamefont {Ferencz}, \citenamefont {Spielmann},
  \citenamefont {Sartania},\ and\ \citenamefont {Krausz}}]{NisoliOL1997}%
  \BibitemOpen
  \bibfield  {author} {\bibinfo {author} {\bibfnamefont {M.}~\bibnamefont
  {Nisoli}}, \bibinfo {author} {\bibfnamefont {S.~D.}\ \bibnamefont
  {Silvestri}}, \bibinfo {author} {\bibfnamefont {O.}~\bibnamefont {Svelto}},
  \bibinfo {author} {\bibfnamefont {R.}~\bibnamefont {Szip\"{o}cs}}, \bibinfo
  {author} {\bibfnamefont {K.}~\bibnamefont {Ferencz}}, \bibinfo {author}
  {\bibfnamefont {C.}~\bibnamefont {Spielmann}}, \bibinfo {author}
  {\bibfnamefont {S.}~\bibnamefont {Sartania}},\ and\ \bibinfo {author}
  {\bibfnamefont {F.}~\bibnamefont {Krausz}},\ }\bibfield  {title} {\bibinfo
  {title} {Compression of high-energy laser pulses below 5 fs},\ }\href@noop {}
  {\bibfield  {journal} {\bibinfo  {journal} {Opt. Lett.}\ }\textbf {\bibinfo
  {volume} {22}},\ \bibinfo {pages} {522} (\bibinfo {year} {1997})}\BibitemShut
  {NoStop}%
\bibitem [{\citenamefont {Lu}\ \emph {et~al.}(2014)\citenamefont {Lu},
  \citenamefont {Tsou}, \citenamefont {Chen}, \citenamefont {Chen},
  \citenamefont {Cheng}, \citenamefont {Yang}, \citenamefont {Chen},
  \citenamefont {Hsu},\ and\ \citenamefont {Kung}}]{LuOptica2014}%
  \BibitemOpen
  \bibfield  {author} {\bibinfo {author} {\bibfnamefont {C.}~\bibnamefont
  {Lu}}, \bibinfo {author} {\bibfnamefont {Y.}~\bibnamefont {Tsou}}, \bibinfo
  {author} {\bibfnamefont {H.}~\bibnamefont {Chen}}, \bibinfo {author}
  {\bibfnamefont {B.}~\bibnamefont {Chen}}, \bibinfo {author} {\bibfnamefont
  {Y.}~\bibnamefont {Cheng}}, \bibinfo {author} {\bibfnamefont
  {S.}~\bibnamefont {Yang}}, \bibinfo {author} {\bibfnamefont {M.}~\bibnamefont
  {Chen}}, \bibinfo {author} {\bibfnamefont {C.}~\bibnamefont {Hsu}},\ and\
  \bibinfo {author} {\bibfnamefont {A.}~\bibnamefont {Kung}},\ }\bibfield
  {title} {\bibinfo {title} {Generation of intense supercontinuum in condensed
  media},\ }\href@noop {} {\bibfield  {journal} {\bibinfo  {journal} {Optica}\
  }\textbf {\bibinfo {volume} {1}},\ \bibinfo {pages} {400} (\bibinfo {year}
  {2014})}\BibitemShut {NoStop}%
\bibitem [{\citenamefont {Vampa}\ \emph {et~al.}(2018)\citenamefont {Vampa},
  \citenamefont {Hammond}, \citenamefont {Nesrallah}, \citenamefont {Naumov},
  \citenamefont {Corkum},\ and\ \citenamefont {Brabec}}]{VampaScience2018}%
  \BibitemOpen
  \bibfield  {author} {\bibinfo {author} {\bibfnamefont {G.}~\bibnamefont
  {Vampa}}, \bibinfo {author} {\bibfnamefont {T.}~\bibnamefont {Hammond}},
  \bibinfo {author} {\bibfnamefont {M.}~\bibnamefont {Nesrallah}}, \bibinfo
  {author} {\bibfnamefont {A.~Y.}\ \bibnamefont {Naumov}}, \bibinfo {author}
  {\bibfnamefont {P.~B.}\ \bibnamefont {Corkum}},\ and\ \bibinfo {author}
  {\bibfnamefont {T.}~\bibnamefont {Brabec}},\ }\bibfield  {title} {\bibinfo
  {title} {Light amplification by seeded kerr instability},\ }\href@noop {}
  {\bibfield  {journal} {\bibinfo  {journal} {Science}\ }\textbf {\bibinfo
  {volume} {359}},\ \bibinfo {pages} {673} (\bibinfo {year}
  {2018})}\BibitemShut {NoStop}%
\bibitem [{\citenamefont {Arachchige}\ \emph {et~al.}(2021)\citenamefont
  {Arachchige}, \citenamefont {Stephen},\ and\ \citenamefont
  {Hammond}}]{JACOL2021}%
  \BibitemOpen
  \bibfield  {author} {\bibinfo {author} {\bibfnamefont {C.}~\bibnamefont
  {Arachchige}}, \bibinfo {author} {\bibfnamefont {J.}~\bibnamefont
  {Stephen}},\ and\ \bibinfo {author} {\bibfnamefont {T.}~\bibnamefont
  {Hammond}},\ }\bibfield  {title} {\bibinfo {title} {Amplification of
  femtosecond pulses based on $\chi^{(3)}$ nonlinear susceptibility in mgo},\
  }\href@noop {} {\bibfield  {journal} {\bibinfo  {journal} {Opt. Lett.}\
  }\textbf {\bibinfo {volume} {46}},\ \bibinfo {pages} {5521} (\bibinfo {year}
  {2021})}\BibitemShut {NoStop}%
\bibitem [{\citenamefont {Valtna}\ \emph {et~al.}(2008)\citenamefont {Valtna},
  \citenamefont {Tamo\v{s}auskas}, \citenamefont {Dubietis},\ and\
  \citenamefont {Piskarskas}}]{ValtnaOL2008}%
  \BibitemOpen
  \bibfield  {author} {\bibinfo {author} {\bibfnamefont {H.}~\bibnamefont
  {Valtna}}, \bibinfo {author} {\bibfnamefont {G.}~\bibnamefont
  {Tamo\v{s}auskas}}, \bibinfo {author} {\bibfnamefont {A.}~\bibnamefont
  {Dubietis}},\ and\ \bibinfo {author} {\bibfnamefont {A.}~\bibnamefont
  {Piskarskas}},\ }\bibfield  {title} {\bibinfo {title} {High-energy broadband
  four-wave optical parametric amplification in bulk fused silica},\
  }\href@noop {} {\bibfield  {journal} {\bibinfo  {journal} {Opt Lett}\
  }\textbf {\bibinfo {volume} {33}},\ \bibinfo {pages} {971} (\bibinfo {year}
  {2008})}\BibitemShut {NoStop}%
\bibitem [{\citenamefont {Nesrallah}\ \emph {et~al.}(2018)\citenamefont
  {Nesrallah}, \citenamefont {Vampa}, \citenamefont {Bart}, \citenamefont
  {Corkum}, \citenamefont {McDonald},\ and\ \citenamefont
  {Brabec}}]{NesrallahOptica2018}%
  \BibitemOpen
  \bibfield  {author} {\bibinfo {author} {\bibfnamefont {M.}~\bibnamefont
  {Nesrallah}}, \bibinfo {author} {\bibfnamefont {G.}~\bibnamefont {Vampa}},
  \bibinfo {author} {\bibfnamefont {G.}~\bibnamefont {Bart}}, \bibinfo {author}
  {\bibfnamefont {P.~B.}\ \bibnamefont {Corkum}}, \bibinfo {author}
  {\bibfnamefont {C.~R.}\ \bibnamefont {McDonald}},\ and\ \bibinfo {author}
  {\bibfnamefont {T.}~\bibnamefont {Brabec}},\ }\bibfield  {title} {\bibinfo
  {title} {Theory of kerr instability amplification},\ }\href@noop {}
  {\bibfield  {journal} {\bibinfo  {journal} {Optica}\ }\textbf {\bibinfo
  {volume} {5}},\ \bibinfo {pages} {271} (\bibinfo {year} {2018})}\BibitemShut
  {NoStop}%
\bibitem [{\citenamefont {Nesrallah}\ \emph {et~al.}(2021)\citenamefont
  {Nesrallah}, \citenamefont {Hammond}, \citenamefont {Hakami}, \citenamefont
  {Bart}, \citenamefont {McDonald}, \citenamefont {Brabec},\ and\ \citenamefont
  {Vampa}}]{Nesrallahbook}%
  \BibitemOpen
  \bibfield  {author} {\bibinfo {author} {\bibfnamefont {M.}~\bibnamefont
  {Nesrallah}}, \bibinfo {author} {\bibfnamefont {T.}~\bibnamefont {Hammond}},
  \bibinfo {author} {\bibfnamefont {A.}~\bibnamefont {Hakami}}, \bibinfo
  {author} {\bibfnamefont {G.}~\bibnamefont {Bart}}, \bibinfo {author}
  {\bibfnamefont {C.}~\bibnamefont {McDonald}}, \bibinfo {author}
  {\bibfnamefont {T.}~\bibnamefont {Brabec}},\ and\ \bibinfo {author}
  {\bibfnamefont {G.}~\bibnamefont {Vampa}},\ }\bibfield  {title} {\bibinfo
  {title} {Kerr instability amplification},\ }in\ \href@noop {} {\emph
  {\bibinfo {booktitle} {Light filaments: structures, challenges, and
  applications}}},\ \bibinfo {editor} {edited by\ \bibinfo {editor}
  {\bibfnamefont {J.}~\bibnamefont {Diels}}, \bibinfo {editor} {\bibfnamefont
  {M.}~\bibnamefont {Richardson}},\ and\ \bibinfo {editor} {\bibfnamefont
  {L.}~\bibnamefont {Arissian}}}\ (\bibinfo  {publisher} {Institution of
  engineering and technology},\ \bibinfo {address} {United Kingdom},\ \bibinfo
  {year} {2021})\ Chap.~\bibinfo {chapter} {10}, pp.\ \bibinfo {pages} {241 --
  260}\BibitemShut {NoStop}%
\bibitem [{\citenamefont {Ghosh}\ \emph {et~al.}()\citenamefont {Ghosh},
  \citenamefont {Drouillard},\ and\ \citenamefont {Hammond}}]{GhoshXXX}%
  \BibitemOpen
  \bibfield  {author} {\bibinfo {author} {\bibfnamefont {S.}~\bibnamefont
  {Ghosh}}, \bibinfo {author} {\bibfnamefont {N.}~\bibnamefont {Drouillard}},\
  and\ \bibinfo {author} {\bibfnamefont {T.}~\bibnamefont {Hammond}},\
  }\bibfield  {title} {\bibinfo {title} {Supercontinuum amplification by kerr
  instability},\ }\href@noop {} {\bibinfo  {journal} {submitted}\ }\BibitemShut
  {NoStop}%
\bibitem [{\citenamefont {Bloembergen}(1980)}]{BloembergenJOSA1980}%
  \BibitemOpen
\bibfield  {journal} {  }\bibfield  {author} {\bibinfo {author} {\bibfnamefont
  {N.}~\bibnamefont {Bloembergen}},\ }\bibfield  {title} {\bibinfo {title}
  {Conservation laws in nonlinear optics},\ }\href@noop {} {\bibfield
  {journal} {\bibinfo  {journal} {J. Opt. Soc. Am.}\ }\textbf {\bibinfo
  {volume} {70}},\ \bibinfo {pages} {1429} (\bibinfo {year}
  {1980})}\BibitemShut {NoStop}%
\bibitem [{\citenamefont {Dubietis}\ \emph {et~al.}(2008)\citenamefont
  {Dubietis}, \citenamefont {Tamo\v{s}auskas}, \citenamefont {Valiulis},\ and\
  \citenamefont {Piskarskas}}]{DubietisLC2008}%
  \BibitemOpen
  \bibfield  {author} {\bibinfo {author} {\bibfnamefont {A.}~\bibnamefont
  {Dubietis}}, \bibinfo {author} {\bibfnamefont {G.}~\bibnamefont
  {Tamo\v{s}auskas}}, \bibinfo {author} {\bibfnamefont {G.}~\bibnamefont
  {Valiulis}},\ and\ \bibinfo {author} {\bibfnamefont {A.}~\bibnamefont
  {Piskarskas}},\ }\bibfield  {title} {\bibinfo {title} {Ultrafast four-wave
  optical parametric amplification in transparent condensed bulk media},\
  }\href@noop {} {\bibfield  {journal} {\bibinfo  {journal} {Laser Chem.}\
  }\textbf {\bibinfo {volume} {2008}},\ \bibinfo {pages} {534951} (\bibinfo
  {year} {2008})}\BibitemShut {NoStop}%
\bibitem [{\citenamefont {Kobayashi}\ \emph
  {et~al.}(2012{\natexlab{b}})\citenamefont {Kobayashi}, \citenamefont {Liu},\
  and\ \citenamefont {Kida}}]{KobayashiIEEE2012}%
  \BibitemOpen
  \bibfield  {author} {\bibinfo {author} {\bibfnamefont {T.}~\bibnamefont
  {Kobayashi}}, \bibinfo {author} {\bibfnamefont {J.}~\bibnamefont {Liu}},\
  and\ \bibinfo {author} {\bibfnamefont {Y.}~\bibnamefont {Kida}},\ }\bibfield
  {title} {\bibinfo {title} {Generation and optimization of femtosecond pulses
  by four-wave mixing process},\ }\href@noop {} {\bibfield  {journal} {\bibinfo
   {journal} {IEEE J. Sel. Top. Quantum Electron.}\ }\textbf {\bibinfo {volume}
  {18}},\ \bibinfo {pages} {54} (\bibinfo {year}
  {2012}{\natexlab{b}})}\BibitemShut {NoStop}%
\bibitem [{\citenamefont {Rubino}\ \emph {et~al.}(2011)\citenamefont {Rubino},
  \citenamefont {Darginavi\v{c}ius}, \citenamefont {Faccio}, \citenamefont
  {Trapani}, \citenamefont {Piskarskas},\ and\ \citenamefont
  {Dubietis}}]{RubinoOL2011}%
  \BibitemOpen
  \bibfield  {author} {\bibinfo {author} {\bibfnamefont {E.}~\bibnamefont
  {Rubino}}, \bibinfo {author} {\bibfnamefont {J.}~\bibnamefont
  {Darginavi\v{c}ius}}, \bibinfo {author} {\bibfnamefont {D.}~\bibnamefont
  {Faccio}}, \bibinfo {author} {\bibfnamefont {P.~D.}\ \bibnamefont {Trapani}},
  \bibinfo {author} {\bibfnamefont {A.}~\bibnamefont {Piskarskas}},\ and\
  \bibinfo {author} {\bibfnamefont {A.}~\bibnamefont {Dubietis}},\ }\bibfield
  {title} {\bibinfo {title} {Generation of broadly tunable sub-30-fs infrared
  pulses by four-wave optical parametric amplification},\ }\href@noop {}
  {\bibfield  {journal} {\bibinfo  {journal} {Opt. Lett.}\ }\textbf {\bibinfo
  {volume} {36}},\ \bibinfo {pages} {382} (\bibinfo {year} {2011})}\BibitemShut
  {NoStop}%
\bibitem [{\citenamefont {You}\ \emph {et~al.}(2016)\citenamefont {You},
  \citenamefont {Reis},\ and\ \citenamefont {Ghimire}}]{YouNatPhys2016}%
  \BibitemOpen
  \bibfield  {author} {\bibinfo {author} {\bibfnamefont {Y.}~\bibnamefont
  {You}}, \bibinfo {author} {\bibfnamefont {D.}~\bibnamefont {Reis}},\ and\
  \bibinfo {author} {\bibfnamefont {S.}~\bibnamefont {Ghimire}},\ }\bibfield
  {title} {\bibinfo {title} {Anisotropic high-harmonic generation in bulk
  crystals},\ }\href@noop {} {\bibfield  {journal} {\bibinfo  {journal} {Nat.
  Phys.}\ }\textbf {\bibinfo {volume} {13}},\ \bibinfo {pages} {345} (\bibinfo
  {year} {2016})}\BibitemShut {NoStop}%
\bibitem [{\citenamefont {Korobenko}\ \emph {et~al.}(2019)\citenamefont
  {Korobenko}, \citenamefont {Hammond}, \citenamefont {Zhang}, \citenamefont
  {Naumov}, \citenamefont {Villeneuve},\ and\ \citenamefont
  {Corkum}}]{KorobenkoOptExp2019}%
  \BibitemOpen
  \bibfield  {author} {\bibinfo {author} {\bibfnamefont {A.}~\bibnamefont
  {Korobenko}}, \bibinfo {author} {\bibfnamefont {T.}~\bibnamefont {Hammond}},
  \bibinfo {author} {\bibfnamefont {C.}~\bibnamefont {Zhang}}, \bibinfo
  {author} {\bibfnamefont {A.~Y.}\ \bibnamefont {Naumov}}, \bibinfo {author}
  {\bibfnamefont {D.~M.}\ \bibnamefont {Villeneuve}},\ and\ \bibinfo {author}
  {\bibfnamefont {P.~B.}\ \bibnamefont {Corkum}},\ }\bibfield  {title}
  {\bibinfo {title} {High-harmonic generation in solids driven by
  counter-propagating pulses},\ }\href@noop {} {\bibfield  {journal} {\bibinfo
  {journal} {Opt. Express}\ }\textbf {\bibinfo {volume} {27}},\ \bibinfo
  {pages} {32630} (\bibinfo {year} {2019})}\BibitemShut {NoStop}%
\bibitem [{\citenamefont {Ko}\ \emph {et~al.}(2020)\citenamefont {Ko},
  \citenamefont {Brown}, \citenamefont {Zhang},\ and\ \citenamefont
  {Corkum}}]{KoJPB2020}%
  \BibitemOpen
  \bibfield  {author} {\bibinfo {author} {\bibfnamefont {D.~H.}\ \bibnamefont
  {Ko}}, \bibinfo {author} {\bibfnamefont {G.~G.}\ \bibnamefont {Brown}},
  \bibinfo {author} {\bibfnamefont {C.}~\bibnamefont {Zhang}},\ and\ \bibinfo
  {author} {\bibfnamefont {P.~B.}\ \bibnamefont {Corkum}},\ }\bibfield  {title}
  {\bibinfo {title} {Delay measurements of attosecond emission in solids},\
  }\href@noop {} {\bibfield  {journal} {\bibinfo  {journal} {J. Phys. B: At.
  Mol. and Opt. Phys.}\ }\textbf {\bibinfo {volume} {53}},\ \bibinfo {pages}
  {124001} (\bibinfo {year} {2020})}\BibitemShut {NoStop}%
\bibitem [{\citenamefont {Alam}\ \emph {et~al.}(2016)\citenamefont {Alam},
  \citenamefont {Leon},\ and\ \citenamefont {Boyd}}]{AlamScience2016}%
  \BibitemOpen
  \bibfield  {author} {\bibinfo {author} {\bibfnamefont {M.}~\bibnamefont
  {Alam}}, \bibinfo {author} {\bibfnamefont {I.~D.}\ \bibnamefont {Leon}},\
  and\ \bibinfo {author} {\bibfnamefont {R.}~\bibnamefont {Boyd}},\ }\bibfield
  {title} {\bibinfo {title} {Large optical nonlinearity of indium tin oxide in
  its epsilon-near-zero regime},\ }\href@noop {} {\bibfield  {journal}
  {\bibinfo  {journal} {Science}\ }\textbf {\bibinfo {volume} {352}},\ \bibinfo
  {pages} {795} (\bibinfo {year} {2016})}\BibitemShut {NoStop}%
\bibitem [{\citenamefont {Reshef}\ \emph {et~al.}(2019)\citenamefont {Reshef},
  \citenamefont {Leon}, \citenamefont {Alam},\ and\ \citenamefont
  {Boyd}}]{ReshefNatMat2019}%
  \BibitemOpen
  \bibfield  {author} {\bibinfo {author} {\bibfnamefont {O.}~\bibnamefont
  {Reshef}}, \bibinfo {author} {\bibfnamefont {I.~D.}\ \bibnamefont {Leon}},
  \bibinfo {author} {\bibfnamefont {M.}~\bibnamefont {Alam}},\ and\ \bibinfo
  {author} {\bibfnamefont {R.}~\bibnamefont {Boyd}},\ }\bibfield  {title}
  {\bibinfo {title} {Nonlinear optical effects in epsilon-near-zero media},\
  }\href@noop {} {\bibfield  {journal} {\bibinfo  {journal} {Nat. Rev. Mater.}\
  }\textbf {\bibinfo {volume} {4}},\ \bibinfo {pages} {535} (\bibinfo {year}
  {2019})}\BibitemShut {NoStop}%
\bibitem [{\citenamefont {Housakou}\ and\ \citenamefont
  {Herrmann}(2001)}]{HousakouPRL2001}%
  \BibitemOpen
  \bibfield  {author} {\bibinfo {author} {\bibfnamefont {A.}~\bibnamefont
  {Housakou}}\ and\ \bibinfo {author} {\bibfnamefont {J.}~\bibnamefont
  {Herrmann}},\ }\bibfield  {title} {\bibinfo {title} {Supercontinuum
  generation of higher-order solitons by fission in photonic crystal fibers},\
  }\href@noop {} {\bibfield  {journal} {\bibinfo  {journal} {Phys. Rev. Lett.}\
  }\textbf {\bibinfo {volume} {87}},\ \bibinfo {pages} {203901} (\bibinfo
  {year} {2001})}\BibitemShut {NoStop}%
\bibitem [{\citenamefont {Berg\'{e}}\ \emph {et~al.}(2013)\citenamefont
  {Berg\'{e}}, \citenamefont {Rolle},\ and\ \citenamefont
  {K\"{o}hler}}]{BergePRA2013}%
  \BibitemOpen
  \bibfield  {author} {\bibinfo {author} {\bibfnamefont {L.}~\bibnamefont
  {Berg\'{e}}}, \bibinfo {author} {\bibfnamefont {J.}~\bibnamefont {Rolle}},\
  and\ \bibinfo {author} {\bibfnamefont {C.}~\bibnamefont {K\"{o}hler}},\
  }\bibfield  {title} {\bibinfo {title} {Enhanced self-compression of
  mid-infrared laser filaments in argon},\ }\href@noop {} {\bibfield  {journal}
  {\bibinfo  {journal} {Phys. Rev. A}\ }\textbf {\bibinfo {volume} {88}},\
  \bibinfo {pages} {023816} (\bibinfo {year} {2013})}\BibitemShut {NoStop}%
\bibitem [{\citenamefont {Hammond}\ \emph {et~al.}(2017)\citenamefont
  {Hammond}, \citenamefont {Villeneuve},\ and\ \citenamefont
  {Corkum}}]{HammondOptica2017}%
  \BibitemOpen
  \bibfield  {author} {\bibinfo {author} {\bibfnamefont {T.}~\bibnamefont
  {Hammond}}, \bibinfo {author} {\bibfnamefont {D.}~\bibnamefont
  {Villeneuve}},\ and\ \bibinfo {author} {\bibfnamefont {P.}~\bibnamefont
  {Corkum}},\ }\bibfield  {title} {\bibinfo {title} {Producing and controlling
  half-cycle near-infrared electric-field transients},\ }\href@noop {}
  {\bibfield  {journal} {\bibinfo  {journal} {Optica}\ }\textbf {\bibinfo
  {volume} {4}},\ \bibinfo {pages} {826} (\bibinfo {year} {2017})}\BibitemShut
  {NoStop}%
\bibitem [{\citenamefont {Agrawal}(2001)}]{AgrawalNFO}%
  \BibitemOpen
  \bibfield  {author} {\bibinfo {author} {\bibfnamefont {G.}~\bibnamefont
  {Agrawal}},\ }\href@noop {} {\emph {\bibinfo {title} {Nonlinear Fiber
  Optics}}},\ \bibinfo {edition} {3rd}\ ed.\ (\bibinfo  {publisher} {Academic
  Press},\ \bibinfo {year} {2001})\BibitemShut {NoStop}%
\bibitem [{\citenamefont {Stephens}\ and\ \citenamefont
  {Malitson}(1952)}]{Stephens1952}%
  \BibitemOpen
  \bibfield  {author} {\bibinfo {author} {\bibfnamefont {R.}~\bibnamefont
  {Stephens}}\ and\ \bibinfo {author} {\bibfnamefont {I.}~\bibnamefont
  {Malitson}},\ }\bibfield  {title} {\bibinfo {title} {Index of refraction of
  magnesium oxide},\ }\href@noop {} {\bibfield  {journal} {\bibinfo  {journal}
  {J. Res. Natl. Bur. Stand.}\ }\textbf {\bibinfo {volume} {49}},\ \bibinfo
  {pages} {249} (\bibinfo {year} {1952})}\BibitemShut {NoStop}%
\bibitem [{\citenamefont {Adair}\ \emph {et~al.}(1989)\citenamefont {Adair},
  \citenamefont {Chase},\ and\ \citenamefont {Payne}}]{AdairPRB1989}%
  \BibitemOpen
  \bibfield  {author} {\bibinfo {author} {\bibfnamefont {R.}~\bibnamefont
  {Adair}}, \bibinfo {author} {\bibfnamefont {L.~L.}\ \bibnamefont {Chase}},\
  and\ \bibinfo {author} {\bibfnamefont {S.~A.}\ \bibnamefont {Payne}},\
  }\bibfield  {title} {\bibinfo {title} {Nonlinear refractive index of optical
  crystals},\ }\href@noop {} {\bibfield  {journal} {\bibinfo  {journal} {Phys.
  Rev. B}\ }\textbf {\bibinfo {volume} {39}},\ \bibinfo {pages} {3337}
  (\bibinfo {year} {1989})}\BibitemShut {NoStop}%
\bibitem [{\citenamefont {Weigand}\ and\ \citenamefont
  {Crespo}(2015)}]{WeigandApplSci2015}%
  \BibitemOpen
  \bibfield  {author} {\bibinfo {author} {\bibfnamefont {R.}~\bibnamefont
  {Weigand}}\ and\ \bibinfo {author} {\bibfnamefont {H.~M.}\ \bibnamefont
  {Crespo}},\ }\bibfield  {title} {\bibinfo {title} {Fundamentals of highly
  non-degenerate cascaded four-wave mixing},\ }\href@noop {} {\bibfield
  {journal} {\bibinfo  {journal} {Appl. Sci}\ }\textbf {\bibinfo {volume}
  {5}},\ \bibinfo {pages} {485} (\bibinfo {year} {2015})}\BibitemShut {NoStop}%
\bibitem [{\citenamefont {B\'{e}jot}\ \emph {et~al.}(2011)\citenamefont
  {B\'{e}jot}, \citenamefont {Kibler}, \citenamefont {Hertz}, \citenamefont
  {Lavorel},\ and\ \citenamefont {Faucher}}]{BejotPRA2011}%
  \BibitemOpen
  \bibfield  {author} {\bibinfo {author} {\bibfnamefont {P.}~\bibnamefont
  {B\'{e}jot}}, \bibinfo {author} {\bibfnamefont {B.}~\bibnamefont {Kibler}},
  \bibinfo {author} {\bibfnamefont {E.}~\bibnamefont {Hertz}}, \bibinfo
  {author} {\bibfnamefont {B.}~\bibnamefont {Lavorel}},\ and\ \bibinfo {author}
  {\bibfnamefont {O.}~\bibnamefont {Faucher}},\ }\bibfield  {title} {\bibinfo
  {title} {General approach to spatiotemporal modulational instability
  processes},\ }\href@noop {} {\bibfield  {journal} {\bibinfo  {journal} {Phys.
  Rev. A}\ }\textbf {\bibinfo {volume} {83}},\ \bibinfo {pages} {013830}
  (\bibinfo {year} {2011})}\BibitemShut {NoStop}%
\bibitem [{\citenamefont {Xu}\ \emph {et~al.}(2006)\citenamefont {Xu},
  \citenamefont {Jia}, \citenamefont {Sun}, \citenamefont {Li}, \citenamefont
  {Li}, \citenamefont {Feng}, \citenamefont {Qiu},\ and\ \citenamefont
  {Xu}}]{XuOptComm2006}%
  \BibitemOpen
  \bibfield  {author} {\bibinfo {author} {\bibfnamefont {S.}~\bibnamefont
  {Xu}}, \bibinfo {author} {\bibfnamefont {T.}~\bibnamefont {Jia}}, \bibinfo
  {author} {\bibfnamefont {H.}~\bibnamefont {Sun}}, \bibinfo {author}
  {\bibfnamefont {C.}~\bibnamefont {Li}}, \bibinfo {author} {\bibfnamefont
  {X.}~\bibnamefont {Li}}, \bibinfo {author} {\bibfnamefont {D.}~\bibnamefont
  {Feng}}, \bibinfo {author} {\bibfnamefont {J.}~\bibnamefont {Qiu}},\ and\
  \bibinfo {author} {\bibfnamefont {Z.}~\bibnamefont {Xu}},\ }\bibfield
  {title} {\bibinfo {title} {Mechanisms of femtosecond laser-induced breakdown
  and damage in mgo},\ }\href@noop {} {\bibfield  {journal} {\bibinfo
  {journal} {Opt. Commun.}\ }\textbf {\bibinfo {volume} {259}},\ \bibinfo
  {pages} {274} (\bibinfo {year} {2006})}\BibitemShut {NoStop}%
\bibitem [{\citenamefont {Backus}\ \emph {et~al.}(1998)\citenamefont {Backus},
  \citenamefont {Durfee}, \citenamefont {Murnane},\ and\ \citenamefont
  {Kapteyn}}]{BackusRSI1998}%
  \BibitemOpen
  \bibfield  {author} {\bibinfo {author} {\bibfnamefont {S.}~\bibnamefont
  {Backus}}, \bibinfo {author} {\bibfnamefont {C.}~\bibnamefont {Durfee}},
  \bibinfo {author} {\bibfnamefont {M.}~\bibnamefont {Murnane}},\ and\ \bibinfo
  {author} {\bibfnamefont {H.}~\bibnamefont {Kapteyn}},\ }\bibfield  {title}
  {\bibinfo {title} {High power ultrafast lasers},\ }\href@noop {} {\bibfield
  {journal} {\bibinfo  {journal} {Rev. Sci. Instrum.}\ }\textbf {\bibinfo
  {volume} {69}},\ \bibinfo {pages} {1207} (\bibinfo {year}
  {1998})}\BibitemShut {NoStop}%
\bibitem [{\citenamefont {Backus}\ \emph {et~al.}(2017)\citenamefont {Backus},
  \citenamefont {Kirchner}, \citenamefont {Lemons}, \citenamefont {Schmidt},
  \citenamefont {Durfee}, \citenamefont {Murnane},\ and\ \citenamefont
  {Kapteyn}}]{BackusOptExp2017}%
  \BibitemOpen
  \bibfield  {author} {\bibinfo {author} {\bibfnamefont {S.}~\bibnamefont
  {Backus}}, \bibinfo {author} {\bibfnamefont {M.}~\bibnamefont {Kirchner}},
  \bibinfo {author} {\bibfnamefont {R.}~\bibnamefont {Lemons}}, \bibinfo
  {author} {\bibfnamefont {D.}~\bibnamefont {Schmidt}}, \bibinfo {author}
  {\bibfnamefont {C.}~\bibnamefont {Durfee}}, \bibinfo {author} {\bibfnamefont
  {M.}~\bibnamefont {Murnane}},\ and\ \bibinfo {author} {\bibfnamefont
  {H.}~\bibnamefont {Kapteyn}},\ }\bibfield  {title} {\bibinfo {title} {Direct
  diode pumped ti:sapphire ultrafast regenerative amplifier system},\
  }\href@noop {} {\bibfield  {journal} {\bibinfo  {journal} {Opt. Express}\
  }\textbf {\bibinfo {volume} {25}},\ \bibinfo {pages} {3666} (\bibinfo {year}
  {2017})}\BibitemShut {NoStop}%
\bibitem [{\citenamefont {Wang}\ \emph {et~al.}(2019)\citenamefont {Wang},
  \citenamefont {Shen}, \citenamefont {Zeng}, \citenamefont {Liu},
  \citenamefont {Li},\ and\ \citenamefont {Xu}}]{WangOL2019}%
  \BibitemOpen
  \bibfield  {author} {\bibinfo {author} {\bibfnamefont {P.}~\bibnamefont
  {Wang}}, \bibinfo {author} {\bibfnamefont {X.}~\bibnamefont {Shen}}, \bibinfo
  {author} {\bibfnamefont {Z.}~\bibnamefont {Zeng}}, \bibinfo {author}
  {\bibfnamefont {J.}~\bibnamefont {Liu}}, \bibinfo {author} {\bibfnamefont
  {R.}~\bibnamefont {Li}},\ and\ \bibinfo {author} {\bibfnamefont
  {Z.}~\bibnamefont {Xu}},\ }\bibfield  {title} {\bibinfo {title}
  {High-performance seed pulses at 910 nm for 100 pw laser facilities by using
  single-stage nondegenerate four-wave mixing},\ }\href@noop {} {\bibfield
  {journal} {\bibinfo  {journal} {Opt. Lett.}\ }\textbf {\bibinfo {volume}
  {44}},\ \bibinfo {pages} {3952 } (\bibinfo {year} {2019})}\BibitemShut
  {NoStop}%
\bibitem [{\citenamefont {Fuji}\ and\ \citenamefont
  {Suzuki}(2007)}]{FujiOL2007}%
  \BibitemOpen
  \bibfield  {author} {\bibinfo {author} {\bibfnamefont {T.}~\bibnamefont
  {Fuji}}\ and\ \bibinfo {author} {\bibfnamefont {T.}~\bibnamefont {Suzuki}},\
  }\bibfield  {title} {\bibinfo {title} {Generation of sub-two-cycle
  mid-infrared pulses by four-wave mixing through filamentation in air},\
  }\href@noop {} {\bibfield  {journal} {\bibinfo  {journal} {Opt. Lett.}\
  }\textbf {\bibinfo {volume} {32}},\ \bibinfo {pages} {3330} (\bibinfo {year}
  {2007})}\BibitemShut {NoStop}%
\bibitem [{\citenamefont {Cheng}\ \emph {et~al.}(2012)\citenamefont {Cheng},
  \citenamefont {Reynolds}, \citenamefont {Widgren},\ and\ \citenamefont
  {Khalil}}]{ChengOL2012}%
  \BibitemOpen
  \bibfield  {author} {\bibinfo {author} {\bibfnamefont {M.}~\bibnamefont
  {Cheng}}, \bibinfo {author} {\bibfnamefont {A.}~\bibnamefont {Reynolds}},
  \bibinfo {author} {\bibfnamefont {H.}~\bibnamefont {Widgren}},\ and\ \bibinfo
  {author} {\bibfnamefont {M.}~\bibnamefont {Khalil}},\ }\bibfield  {title}
  {\bibinfo {title} {Generation of tunable octave-spanning mid-infrared pulses
  by filamentation in gas media},\ }\href@noop {} {\bibfield  {journal}
  {\bibinfo  {journal} {Opt. Lett.}\ }\textbf {\bibinfo {volume} {37}},\
  \bibinfo {pages} {1787} (\bibinfo {year} {2012})}\BibitemShut {NoStop}%
\bibitem [{\citenamefont {Fuji}\ \emph {et~al.}(2017)\citenamefont {Fuji},
  \citenamefont {Shirai},\ and\ \citenamefont {Nomura}}]{FujiApplSci2017}%
  \BibitemOpen
  \bibfield  {author} {\bibinfo {author} {\bibfnamefont {T.}~\bibnamefont
  {Fuji}}, \bibinfo {author} {\bibfnamefont {H.}~\bibnamefont {Shirai}},\ and\
  \bibinfo {author} {\bibfnamefont {Y.}~\bibnamefont {Nomura}},\ }\bibfield
  {title} {\bibinfo {title} {Development and application of sub-cycle
  mid-infrared source based on laser filamentation},\ }\href@noop {} {\bibfield
   {journal} {\bibinfo  {journal} {Appl. Sci.}\ }\textbf {\bibinfo {volume}
  {7}},\ \bibinfo {pages} {857} (\bibinfo {year} {2017})}\BibitemShut {NoStop}%
\bibitem [{\citenamefont {Voronin}\ \emph {et~al.}(2014)\citenamefont
  {Voronin}, \citenamefont {Nomura}, \citenamefont {Shirai}, \citenamefont
  {Fuji},\ and\ \citenamefont {Zheltikov}}]{VoroninAPB2014}%
  \BibitemOpen
  \bibfield  {author} {\bibinfo {author} {\bibfnamefont {A.}~\bibnamefont
  {Voronin}}, \bibinfo {author} {\bibfnamefont {Y.}~\bibnamefont {Nomura}},
  \bibinfo {author} {\bibfnamefont {H.}~\bibnamefont {Shirai}}, \bibinfo
  {author} {\bibfnamefont {T.}~\bibnamefont {Fuji}},\ and\ \bibinfo {author}
  {\bibfnamefont {A.}~\bibnamefont {Zheltikov}},\ }\bibfield  {title} {\bibinfo
  {title} {Half-cycle pulses in the mid-infrared from a two-color laser-induced
  filament},\ }\href@noop {} {\bibfield  {journal} {\bibinfo  {journal} {Appl.
  Phys. B}\ }\textbf {\bibinfo {volume} {117}},\ \bibinfo {pages} {611}
  (\bibinfo {year} {2014})}\BibitemShut {NoStop}%
\bibitem [{\citenamefont {Shirai}\ \emph {et~al.}(2018)\citenamefont {Shirai},
  \citenamefont {Kumaki}, \citenamefont {Nomura},\ and\ \citenamefont
  {Fuji}}]{ShiraiOL2018}%
  \BibitemOpen
  \bibfield  {author} {\bibinfo {author} {\bibfnamefont {H.}~\bibnamefont
  {Shirai}}, \bibinfo {author} {\bibfnamefont {F.}~\bibnamefont {Kumaki}},
  \bibinfo {author} {\bibfnamefont {Y.}~\bibnamefont {Nomura}},\ and\ \bibinfo
  {author} {\bibfnamefont {T.}~\bibnamefont {Fuji}},\ }\bibfield  {title}
  {\bibinfo {title} {High-harmonic generation in solids driven by subcycle
  midinfrared pulses from two-color filamentation},\ }\href@noop {} {\bibfield
  {journal} {\bibinfo  {journal} {Opt. Lett.}\ }\textbf {\bibinfo {volume}
  {43}},\ \bibinfo {pages} {2094} (\bibinfo {year} {2018})}\BibitemShut
  {NoStop}%
\bibitem [{\citenamefont {He}\ \emph {et~al.}(2014)\citenamefont {He},
  \citenamefont {Liu},\ and\ \citenamefont {Kobayashi}}]{HeApplSci2014}%
  \BibitemOpen
  \bibfield  {author} {\bibinfo {author} {\bibfnamefont {J.}~\bibnamefont
  {He}}, \bibinfo {author} {\bibfnamefont {J.}~\bibnamefont {Liu}},\ and\
  \bibinfo {author} {\bibfnamefont {T.}~\bibnamefont {Kobayashi}},\ }\bibfield
  {title} {\bibinfo {title} {Tunable multicolored femtosecond pulse generation
  using cascaded four-wave mixing in bulk materials},\ }\href@noop {}
  {\bibfield  {journal} {\bibinfo  {journal} {Appl. Sci.}\ }\textbf {\bibinfo
  {volume} {4}},\ \bibinfo {pages} {444} (\bibinfo {year} {2014})}\BibitemShut
  {NoStop}%
\end{thebibliography}

%

\end{document}